\documentclass{aa}
\usepackage{graphicx}
\usepackage{amssymb,amsmath}
\usepackage{latexsym}
\usepackage{mathtools}
\usepackage{epsfig}
\usepackage{natbib}
\usepackage{txfonts}
\begin{document}
   \title{The past and future obliquity of Saturn as Titan migrates}
%
   \author{Melaine Saillenfest\inst{1}
          \and
          Giacomo Lari\inst{2}
          \and
          Gwena{\"e}l Bou{\'e}\inst{1}
          \and
          Ariane Courtot\inst{1}
          }
   \authorrunning{Saillenfest et al.}
   \institute{IMCCE, Observatoire de Paris, PSL Research University, CNRS, Sorbonne Universit\'e, Universit\'e de Lille, 75014 Paris, France
              \email{melaine.saillenfest@obspm.fr}
              \and
              Department of Mathematics, University of Pisa, Largo Bruno Pontecorvo 5, 56127 Pisa, Italy
             }
   \date{Received 12 November 2020 / Accepted 23 January 2021}


  \abstract
  {Giant planets are expected to form with near-zero obliquities. It has recently been shown that the fast migration of Titan could be responsible for the current $26.7^\circ$-tilt of Saturn's spin axis.}
  {We aim to quantify the level of generality of this result by measuring the range of parameters allowing for this scenario to happen. Since Titan continues to migrate today, we also aim to determine the obliquity that Saturn will reach in the future.}
  {For a large variety of migration rates for Titan, we numerically propagated the orientation of Saturn's spin axis both backwards and forwards in time. We explored a broad range of initial conditions after the late planetary migration, including both small and large obliquity values.}
  {In the adiabatic regime, the likelihood of reproducing Saturn's current spin-axis orientation is maximised for primordial obliquities between about $2^\circ$ and $7^\circ$. For a slightly faster migration than expected from radio-science experiments, non-adiabatic effects even allow for exactly null primordial obliquities. Starting from such small tilts, Saturn's spin axis can evolve up to its current state provided that: \emph{i)} the semi-major axis of Titan changed by more than $5\%$ of its current value since the late planetary migration, and \emph{ii)} its migration rate does not exceed ten times the nominal measured rate. In comparison, observational data suggest that the increase in Titan's semi-major axis exceeded $50\%$ over $4$~Gyrs, and error bars imply that the current migration rate is unlikely to be larger than $1.5$ times its nominal value.}
  {If Titan did migrate substantially before today, tilting Saturn from a small obliquity is not only possible, but it is the most likely scenario. Saturn's obliquity is still expected to be increasing today and could exceed $65^\circ$ in the future. Maximising the likelihood would also put strict constraints on Saturn's polar moment of inertia. However, the possibility remains that Saturn's primordial obliquity was already large, for instance as a result of a massive collision. The unambiguous distinction between these two scenarios would be given by a precise measure of Saturn's polar moment of inertia.}


  \maketitle

\section{Introduction}
   The obliquity of a planet is the angle between its spin axis and the normal to its orbit. In the protoplanetary disc, giant planets are expected to form with near-zero obliquities \citep{Ward-Hamilton_2004,Rogoszinski-Hamilton_2020a}. After the formation of Saturn, some dynamical mechanism must therefore have tilted its spin axis up to its current obliquity of $26.7^\circ$.
   
   \cite{Ward-Hamilton_2004} showed that Saturn is currently located very close to a secular spin-orbit resonance with the nodal precession mode of Neptune. This resonance strongly affects Saturn's spin axis today, and it offers a tempting explanation for its current large obliquity. For years, the scenarios that were most successful in reproducing Saturn's current obliquity through this resonance invoked the late planetary migration \citep{Hamilton-Ward_2004,Boue-etal_2009,Vokrouhlicky-Nesvorny_2015,Brasser-Lee_2015}. However, \cite{Saillenfest-etal_2020b} have recently shown that this picture is incompatible with the fast tidal migration of Titan detected by \cite{Lainey-etal_2020} in two independent sets of observations -- assuming that this migration is not specific to the present epoch but went on over a substantial interval of time. Indeed, satellites affect the spin-axis precession of their host planets (see e.g.~\citealp{Ward_1975,Tremaine_1991,Laskar-etal_1993a,Boue-Laskar_2006}). Since the effect of a satellite depends on its orbital distance, migrating satellites induce a long-term drift in the planet's spin-axis precession velocity. In the course of this drift, large obliquity variations can occur if a secular spin-orbit resonance is encountered (i.e. if the planet's spin-axis precession velocity becomes commensurate with a harmonic of its orbital precession). Because of this mechanism, dramatic variations in the Earth's obliquity are expected to take place in a few billion years from now, as a result of the Moon's migration \citep{NerondeSurgy-Laskar_1997}. Likewise, Jupiter's obliquity is likely steadily increasing today and could exceed $30^\circ$ in the next billions of years, as a result of the migration of the Galilean satellites \citep{Lari-etal_2020,Saillenfest-etal_2020}.

   A significant migration of Saturn's satellites implies that, contrary to previous assumptions, Saturn's spin-axis precession velocity was much smaller in the past, precluding any resonance with an orbital frequency. The same conclusion could also hold for Jupiter \citep{Lainey-etal_2009,Lari-etal_2020}. In fact, \cite{Saillenfest-etal_2020b} have shown that Titan's migration itself is likely responsible for the resonant encounter between Saturn's spin axis and the nodal precession mode of Neptune. Their results indicate that this relatively recent resonant encounter could explain the current large obliquity of Saturn starting from a small value, possibly less than $3^\circ$. This new paradigm solves the problem of the low probability of reproducing both the orbits and axis tilts of Jupiter and Saturn during the late planetary migration \citep{Brasser-Lee_2015}. However, it revokes the concomitant constraints on the parameters of the late planetary migration.
   
   The findings of \cite{Saillenfest-etal_2020b} have been obtained through backward integrations from Saturn's current spin orientation, and by exploring migration histories for Titan in the vicinity of the nominal scenario of \cite{Lainey-etal_2020}. However, observation uncertainties and our lack of knowledge about the past evolution of Titan's migration rate still allow for a large variety of migration histories, and one can wonder whether the dramatic influence of Titan is a generic result or whether it is restricted to the range of parameters explored by \cite{Saillenfest-etal_2020b}. Moreover, even though backward numerical integrations do prove that Titan's migration is able to raise Saturn's obliquity, a statistical picture of the possible trajectories that could have been followed is still missing. In this regard, the likelihood of following a given dynamical pathway would be quite valuable, because it could be used as a constraint to the parameters of the model, in the spirit of \cite{Boue-etal_2009}, \cite{Brasser-Lee_2015}, and \cite{Vokrouhlicky-Nesvorny_2015}.
   
   For these reasons, we aim to explore the outcomes given by all conceivable migration timescales for Titan, and to perform a statistical search for Saturn's past obliquity. This will provide the whole region of the parameter space allowing Titan's migration to be responsible for Saturn's large obliquity, with the corresponding probability. Finally, since Titan still migrates today, Saturn's obliquity could suffer from further large variations in the future, in the same way as Jupiter \citep{Saillenfest-etal_2020}. Therefore, we also aim to extend previous analyses to the future dynamics of Saturn's spin axis.
   
   Our article is organised as follows. In Sect.~\ref{sec:dyn}, we recall the dynamical model used by \cite{Saillenfest-etal_2020b} and discuss the range of acceptable values for the physical parameters of Saturn and its satellites. Sections~\ref{sec:backward} and \ref{sec:CIsearch} are dedicated to the past spin-axis dynamics of Saturn: after having explored the parameter space and quantified the importance of non-adiabaticity, we perform Monte Carlo experiments to search for the initial conditions of Saturn's spin axis. In Sect.~\ref{sec:fut}, we present our results about the obliquity values that will be reached by Saturn in the future. Finally, our conclusions are summarised in Sect.~\ref{sec:ccl}.
   
\section{Secular dynamics of the spin axis}\label{sec:dyn}

   \subsection{Equations of motion}\label{ssec:eqmot}
   In the approximation of rigid rotation, the spin-axis dynamics of an oblate planet subject to the lowest-order term of the torque from the Sun is given for instance by \cite{Laskar-Robutel_1993} or \cite{NerondeSurgy-Laskar_1997}. Far from spin-orbit resonances, and due to the weakness of the torque, the long-term evolution of the spin axis is accurately described by the secular Hamiltonian function (i.e. averaged over rotational and orbital motions). This Hamiltonian can be written
   \begin{equation}\label{eq:Hinit}
      \begin{aligned}
         \mathcal{H}(X,-\psi,t) &= -\frac{\alpha}{2}\frac{X^2}{\big(1-e(t)^2\big)^{3/2}} \\
         &- \sqrt{1-X^2}\big(\mathcal{A}(t)\sin\psi + \mathcal{B}(t)\cos\psi\big) \\
         &+ 2X\mathcal{C}(t),
      \end{aligned}
   \end{equation}
   where the conjugate coordinates are $X=\cos\varepsilon$ (cosine of obliquity) and $-\psi$ (minus the precession angle). The Hamiltonian in Eq.~\eqref{eq:Hinit} explicitly depends on time $t$ through the orbital eccentricity $e$ of the planet and through the functions
   \begin{equation}
      \left\{
      \begin{aligned}
         \mathcal{A}(t) &= \frac{2\big(\dot{q}+p\,\mathcal{C}(t)\big)}{\sqrt{1-p^2-q^2}}\,, \\
         \mathcal{B}(t) &= \frac{2\big(\dot{p}-q\,\mathcal{C}(t)\big)}{\sqrt{1-p^2-q^2}} \,,\\
      \end{aligned}
      \right.
      \quad
      \text{and}
      \quad
      \mathcal{C}(t) = q\dot{p}-p\dot{q}\,.
   \end{equation}
   In these expressions, $q=\eta\cos\Omega$ and $p=\eta\sin\Omega$, where $\eta\equiv\sin(I/2)$, and $I$ and $\Omega$ are the orbital inclination and the longitude of ascending node of the planet, respectively. The quantity $\alpha$ is called the precession constant. It depends on the spin rate of the planet and on its mass distribution, through the formula:
   \begin{equation}\label{eq:alpha}
      \alpha = \frac{3}{2}\frac{\mathcal{G}m_\odot}{\omega a^3}\frac{J_2}{\lambda} \,,
   \end{equation}
   where $\mathcal{G}$ is the gravitational constant, $m_\odot$ is the mass of the sun, $\omega$ is the spin rate of the planet, $a$ is its semi-major axis, $J_2$ is its second zonal gravity coefficient, and $\lambda$ is its normalised polar moment of inertia. The parameters $J_2$ and $\lambda$ can be expressed as
   \begin{equation}\label{eq:J2lb}
      J_2 = \frac{2C-A-B}{2MR_\mathrm{eq}^2}
      \quad\text{and}\quad
      \lambda = \frac{C}{MR_\mathrm{eq}^2}\,,
   \end{equation}
   where $A$, $B$, and $C$ are the equatorial and polar moments of inertia of the planet, $M$ is its mass, and $R_\mathrm{eq}$ is its equatorial radius.
   
   The precession rate of the planet is increased if it possesses massive satellites. Far-away satellites increase the torque exerted by the sun on the equatorial bulge of the planet, whereas close-in satellites artificially increase the oblateness and the rotational angular momentum of the planet \citep{Boue-Laskar_2006}. In the close-in regime, an expression for the effective precession constant has been derived by \cite{Ward_1975}. It has been generalised by \cite{French-etal_1993} who included the effect of the non-zero orbital inclinations of the satellites, as they oscillate around their local `Laplace plane' (see e.g. \citealp{Tremaine_2009}). The effective precession constant is obtained by replacing $J_2$ and $\lambda$ in Eq.~\eqref{eq:alpha} by the effective values:
   \begin{equation}\label{eq:J2prime}
      \begin{aligned}
         J_2' &= J_2 + \frac{1}{2}\sum_k\frac{m_k}{M}\frac{a_k^2}{R_\mathrm{eq}^2}\frac{\sin(2\varepsilon-2L_k)}{\sin(2\varepsilon)}\,, \\
         \lambda' &= \lambda + \sum_k\frac{m_k}{M}\frac{a_k^2}{R_\mathrm{eq}^2}\frac{n_k}{\omega}\frac{\sin(\varepsilon-L_k)}{\sin(\varepsilon)}\,,
      \end{aligned}
   \end{equation}
   where $m_k$, $a_k$ and $n_k$ are the mass, the semi-major axis, and the mean motion of the $k$th satellite, $\varepsilon$ is the obliquity of the planet, and $L_k$ is the inclination of the Laplace plane of the $k$th satellite with respect to the planet's equator. For regular satellites, $L_k$ lies between $0$ (close-in satellite) and $\varepsilon$ (far-away satellite). The formulas of \cite{French-etal_1993} given by Eq.~\eqref{eq:J2prime} are valid whatever the distance of the satellites, and they closely match the general precession solutions of \cite{Boue-Laskar_2006}. We can also verify that the small eccentricities of Saturn's major satellites do not contribute substantially to $J_2'$ and $\lambda'$, allowing us to neglect them.
   
   Because of its large mass, Titan is by far the satellite that contributes most to the value of~$\alpha$. Therefore, even though its Laplace plane is not much inclined, taking its inclination into account changes the whole satellites' contribution by several percent\footnote{This point was missed by \cite{French-etal_1993} who only included the inclination contribution of Iapetus.}. \cite{Tremaine_2009} give a closed-form expression for $L_k$ in the regime $m_k\ll M$, where all other satellites $j$ with $a_j<a_k$ are also taken into account. The values obtained for Titan ($L_6\approx 0.62^\circ$) and Iapetus ($L_8\approx 16.03^\circ$) are very close to those found in the quasi-periodic decomposition of their ephemerides (see e.g.~\citealp{Vienne-Duriez_1995}). The inclinations $L_k$ of the other satellites of Saturn do not contribute substantially to the value of~$\alpha$.
   
   Even though the value of $\alpha$ computed using Eq.~\eqref{eq:J2prime} yields an accurate value of the current mean spin-axis precession velocity of Saturn as $\dot{\psi}=\alpha X/(1-e^2)^{3/2}$, it cannot be directly used to propagate the dynamics using the Hamiltonian function in Eq.~\eqref{eq:Hinit}, because $\alpha$ would itself be a function of $X$, which contradicts the Hamiltonian formulation. For this reason, authors generally assume that $\alpha$ only weakly depends on $\varepsilon$, such that the satellite's contributions can be considered to be fixed while $\varepsilon$ varies according to Hamilton's equations of motion (see e.g. \citealp{Ward-Hamilton_2004,Boue-etal_2009,Vokrouhlicky-Nesvorny_2015,Brasser-Lee_2015}). In our case, Titan largely dominates the satellite's contribution, and it is almost in the close-in regime $(L_6\ll\varepsilon)$. We can therefore use the same trick as \cite{Saillenfest-etal_2020b} and replace Eq.~\eqref{eq:J2prime} by
   \begin{equation}\label{eq:J2tilde}
      \tilde{J}_2 = J_2 + \frac{1}{2}\frac{\tilde{m}_6}{M}\frac{a_6^2}{R_\mathrm{eq}^2}\,,
      \quad\text{and}\quad
      \tilde{\lambda} = \lambda + \frac{\tilde{m}_6}{M}\frac{a_6^2}{R_\mathrm{eq}^2}\frac{n_6}{\omega}\,,
   \end{equation}
   where only Titan is considered ($k=6$), in the close-in regime ($L_6=0$), and where its mass $m_6$ has been slightly increased ($\tilde{m}_6\approx 1.04\,m_6$) so as to produce the exact same value of $\alpha$ today using Eq.~\eqref{eq:J2tilde} instead of Eq.~\eqref{eq:J2prime}. This slight increase in Titan's mass has no physical meaning; it is only used here to provide the right connection between $\lambda$ and today's value of $\alpha$. This point is further discussed in Sect.~\ref{ssec:alpha}   

   \subsection{Orbital solution}\label{ssec:orbitsol}
   The Hamiltonian function in Eq.~\eqref{eq:Hinit} depends on the orbit of the planet and on its temporal variations. In order to explore the long-term dynamics of Saturn's spin axis, we need an orbital solution that is valid over billions of years. In the same way as \cite{Saillenfest-etal_2020}, we use the secular solution of \cite{Laskar_1990} expanded in quasi-periodic series:
   \begin{equation}\label{eq:qprep}
      \begin{aligned}
         z = e\exp(i\varpi) &= \sum_k E_k\exp(i\theta_k) \,,\\
         \zeta = \eta\exp(i\Omega) &= \sum_k S_k\exp(i\phi_k)\,,
      \end{aligned}
   \end{equation}
   where $\varpi$ is Saturn's longitude of perihelion. The amplitudes $E_k$ and $S_k$ are real constants, and the angles $\theta_k$ and $\phi_k$ evolve linearly over time $t$ with frequencies $\mu_k$ and $\nu_k$:
   \begin{equation}\label{eq:munu}
      \theta_k(t) = \mu_k\,t + \theta_k^{(0)}
      \hspace{0.5cm}\text{and}\hspace{0.5cm}
      \phi_k(t) = \nu_k\,t + \phi_k^{(0)}\,.
   \end{equation}
   See Appendix~\ref{asec:QPS} for the complete orbital solution of \cite{Laskar_1990}.
   
   The series in Eq.~\eqref{eq:qprep} contain contributions from all the planets of the Solar System. In the integrable approximation, the frequency of each term corresponds to a unique combination of the fundamental frequencies of the system, usually noted $g_j$ and $s_j$. In the limit of small masses, small eccentricities and small inclinations (Lagrange-Laplace secular system), the $z$ series only contains the frequencies $g_j$, while the $\zeta$ series only contains the frequencies $s_j$ (see e.g. \citealp{Murray-Dermott_1999} or \citealp{Laskar-etal_2012}). This is not the case in more realistic situations. Table~\ref{tab:zetashort} shows the combinations of fundamental frequencies identified for the largest terms of Saturn's $\zeta$ series obtained by \cite{Laskar_1990}.
   
   \begin{table}
      \caption{First twenty terms of Saturn's inclination and longitude of ascending node in the J2000 ecliptic and equinox reference frame.}
      \label{tab:zetashort}
      \vspace{-0.7cm}
      \begin{equation*}
         \begin{array}{rcrrr}
         \hline
         \hline
         k & \text{identification}\tablefootmark{*} & \nu_k\ (''\,\text{yr}^{-1}) & S_k\times 10^8 & \phi_k^{(0)}\ (^\text{o}) \\
         \hline   
           1 &          s_5 &   0.00000 & 1377395 & 107.59 \\
           2 &          s_6 & -26.33023 &  785009 & 127.29 \\
           3 &          s_8 &  -0.69189 &   55969 &  23.96 \\
           4 &          s_7 &  -3.00557 &   39101 & 140.33 \\
           5 &  g_5-g_6+s_7 & -26.97744 &    5889 &  43.05 \\
           6 &     2g_6-s_6 &  82.77163 &    3417 & 128.95 \\
           7 &  g_5+g_6-s_6 &  58.80017 &    2003 & 212.90 \\
           8 &     2g_5-s_6 &  34.82788 &    1583 & 294.12 \\
           9 &          s_1 &  -5.61755 &    1373 & 168.70 \\
          10 &          s_4 & -17.74818 &    1269 & 123.28 \\
          11 & -g_5+g_7+s_6 & -27.48935 &    1014 & 218.53 \\
          12 &  g_5-g_7+s_6 & -25.17116 &     958 & 215.94 \\
          13 &  g_5-g_6+s_6 & -50.30212 &     943 & 209.84 \\
          14 &  g_5-g_7+s_7 &  -1.84625 &     943 &  35.32 \\
          15 & -g_5+g_6+s_6 &  -2.35835 &     825 & 225.04 \\
          16 & -g_5+g_7+s_7 &  -4.16482 &     756 &  51.51 \\
          17 &          s_2 &  -7.07963 &     668 & 273.79 \\
          18 & -g_6+g_7+s_7 & -28.13656 &     637 & 314.07 \\
          19 &  g_7-g_8+s_7 &  -0.58033 &     544 &  17.32 \\
          20 &  s_1+\gamma  &  -5.50098 &     490 & 162.89 \\
         \hline
         \end{array}
      \end{equation*}
      \vspace{-0.5cm}
      \tablefoot{Due to the secular resonance $(g_1-g_5)-(s_1-s_2)$, an additional fundamental frequency $\gamma$ appears in term 20 (see \citealp{Laskar_1990}).\\
      \tablefoottext{*}{There are typographical errors in \cite{Laskar_1990} in the identification of the 14th and 16th terms.}
      }
   \end{table}
   
   As explained by \cite{Saillenfest-etal_2019a}, at first order in the amplitudes $S_k$ and $E_k$, secular spin-orbit resonant angles can only be of the form $\sigma_p = \psi+\phi_p$, where $p$ is a given index in the $\zeta$ series. Resonances featuring terms of the $z$ series only appear at third order and beyond. For the giant planets of the Solar System, the existing secular spin-orbit resonances are small and isolated from each other, and only first-order resonances play a substantial role (see e.g. \citealp{Saillenfest-etal_2020}).
   
\begin{figure}
      \centering
      \includegraphics[width=\columnwidth]{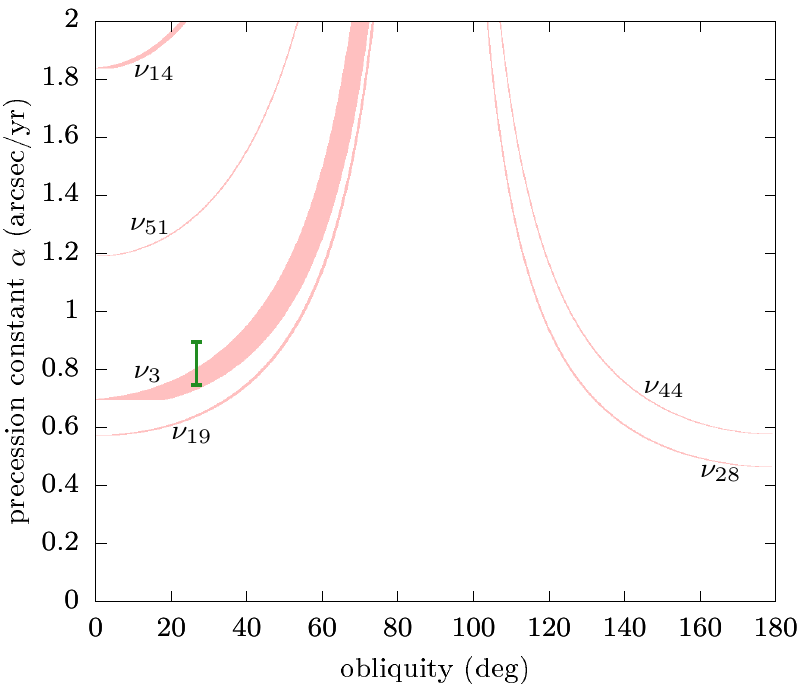}
      \caption{Location and width of every first-order secular spin-orbit resonance for Saturn. Each resonant angle is of the form $\sigma_p = \psi+\phi_p$ where $\phi_p$ has frequency $\nu_p$ labelled on the graph according to its index in the orbital series (see Table~\ref{tab:zetashort} and Appendix~\ref{asec:QPS}). For a given value of the precession constant $\alpha$, the interval of obliquity enclosed by the separatrix is shown in pink, as computed using the formulas of \cite{Saillenfest-etal_2019a}. The green bar shows Saturn's current obliquity and the range for its precession constant considered in this article, as detailed in Sects.~\ref{ssec:alpha} and \ref{ssec:condinit}.}
      \label{fig:widths}
   \end{figure}

   Figure~\ref{fig:widths} shows the location and width of every first-order resonance for the spin-axis of Saturn in an interval of precession constant $\alpha$ ranging from $0$ to $2''\,$yr$^{-1}$. Because of the chaotic dynamics of the Solar System \citep{Laskar_1989}, the fundamental frequencies related to the terrestrial planets (e.g. $s_1$, $s_2$, $s_4$, and $\gamma$ appearing in Table~\ref{tab:zetashort}) could vary substantially over billions of years \citep{Laskar_1990}. However, they only marginally contribute to Saturn's orbital solution and none of them takes part in the resonances shown in Fig.~\ref{fig:widths}. Our secular orbital solution for Saturn can therefore be considered valid since the late planetary migration, which presumably ended at least $4$~Gyrs ago (see e.g. \citealp{Nesvorny-Morbidelli_2012,Deienno_2017,Clement-etal_2018}). For this reason, we consider in all this article a maximum timespan of $4$~Gyrs in the past. As shown by \cite{Saillenfest-etal_2020b}, this timespan is more than enough for Saturn to relax to its primordial obliquity value. Our results are therefore independent of this choice, unless one considers a much slower migration rate for Titan than observed today. This last case is discussed in Sect.~\ref{ssec:adiab}.
   
   \subsection{Precession constant}\label{ssec:alpha}
   As shown by the Hamiltonian function in Eq.~\eqref{eq:Hinit}, the precession constant $\alpha$ is a key parameter of the spin-axis dynamics of a planet. The physical parameters of Saturn that enter into its expression (see Eq.~\ref{eq:alpha}) are all very well constrained from observations, except the normalised polar moment of inertia $\lambda$. Indeed, the gravitational potential measured by spacecrafts only provides differences between the moments of inertia (e.g. the coefficient $J_2$). In order to obtain the individual value of a single moment of inertia, one would need to detect the precession of the spin axis or the Lense--Thirring effect, as explained for instance by \cite{Helled-etal_2011}. Such measurements are difficult considering the limited timespan covered by space missions. To our knowledge, the most accurate estimate of Saturn's polar motion, including decades of astrometric observations and \emph{Cassini} data, is given by \cite{French-etal_2017}. However, their estimate is still not accurate enough to bring any decisive constraint on Saturn's polar moment of inertia. Moreover, since the observed polar motion of Saturn is affected by many short-period harmonics, it cannot be directly linked to the secular spin-axis precession rate $\dot{\psi}$ discussed in this article. Removing short-period harmonics from the observed signal would require an extensive modelling that is not yet available. Even though some attempts to compute a secular trend from Saturn's spin-axis observations have been reported (as the unpublished results of Jacobson cited by \citealp{Vokrouhlicky-Nesvorny_2015}), we must still rely on theoretical values of $\lambda$.
   
   As pointed out by \cite{Saillenfest-etal_2020}, one must be careful about the normalisation used for $\lambda$. Here, we adopt $R_\text{eq}=60268$~km by convention and we renormalise each quantity in Eqs.~\eqref{eq:J2lb} and \eqref{eq:J2prime} accordingly. Many different values of $\lambda$ can be found in the literature. Under basic assumptions, \cite{Jeffreys_1924} obtained a value of $0.198$. This value is smaller than other estimates found in the literature, even though it is marginally compatible with the calculations of \cite{Hubbard-Marley_1989}, who gave $\lambda=0.22037$ with a $10\%$ uncertainty. The latter value and its uncertainty have been reused by many authors afterwards, including \cite{French-etal_1993} and \cite{Ward-Hamilton_2004}. Later on, \cite{Helled-etal_2009} obtained values of $\lambda$ ranging between $0.207$ and $0.210$. From an exploration of the parameter space, \cite{Helled_2011} then found $\lambda\in[0.200,0.205]$, but the normalisation used in this article is ambiguous\footnote{Even though Saturn's mean radius is explicitly mentioned by \cite{Helled_2011}, her values are cited by \cite{Nettelmann-etal_2013} as having been normalised using the equatorial radius instead, according to a `personal communication'.}. The computations of \cite{Nettelmann-etal_2013} yielded yet another range for $\lambda$, estimated to lie in $[0.219,0.220]$. Among the alternative models proposed by \cite{Vazan-etal_2016}, values of $\lambda$ are found to range between $0.222$ and $0.228$. Finally, \cite{Movshovitz-etal_2020} used a new fitting technique supposed to be less model-dependent, and obtained $\lambda\in[0.2204,0.2234]$ at the $3\sigma$ error level (assuming that their values are normalised using $R_\mathrm{eq}$, which is not specified in the article). In the review of \cite{Fortney-etal_2018} focussing on the better knowledge of Saturn's internal structure brought by the \emph{Cassini} mission, the authors go back to a value of $\lambda$ equal to $0.22\pm 10\%$. A value of $0.22$ is also quoted in the review of \cite{Helled_2018}.
   
   Here, instead of relying on one particular estimate of $\lambda$, we turn to the exploration of the whole range of values given in the literature, which is slightly larger than $\lambda\in[0.200,0.240]$. The spin velocity of Saturn is taken from \cite{Archinal-etal_2018} and its $J_2$ from \cite{Iess-etal_2019}. For consistency with Saturn's orbital solution (Sect.~\ref{ssec:orbitsol}), we take its mass and secular semi-major axis from \cite{Laskar_1990}.
   
   In order to compute $J_2'$ and $\lambda'$ in Eq.~\eqref{eq:J2prime}, we need the masses and orbital elements of Saturn's satellites. We take into account the eight major satellites of Saturn and use the masses of the SAT427 numerical ephemerides\footnote{\texttt{https://ssd.jpl.nasa.gov/}}. These ephemerides are then digitally filtered in order to obtain the secular semi-major axes. The inclination $L_k$ of the Laplace plane of each satellite is computed using the formula of \cite{Tremaine_2009}. Taking $\lambda$ into its exploration interval, the current value of Saturn's precession constant, computed from Eqs.~\eqref{eq:alpha} and~\eqref{eq:J2prime}, ranges from $0.747$ to $0.894''\,$yr$^{-1}$. The corresponding adjusted mass of Titan in Eq.~\eqref{eq:J2tilde} is $\tilde{m}_6\approx 1.04\,m_6$. Similar results are obtained when using the more precise values of $L_k$ given by the constant terms of the full series of \cite{Vienne-Duriez_1995} and \cite{Duriez-Vienne_1997}.
   
\begin{figure}
      \centering
      \includegraphics[width=\columnwidth]{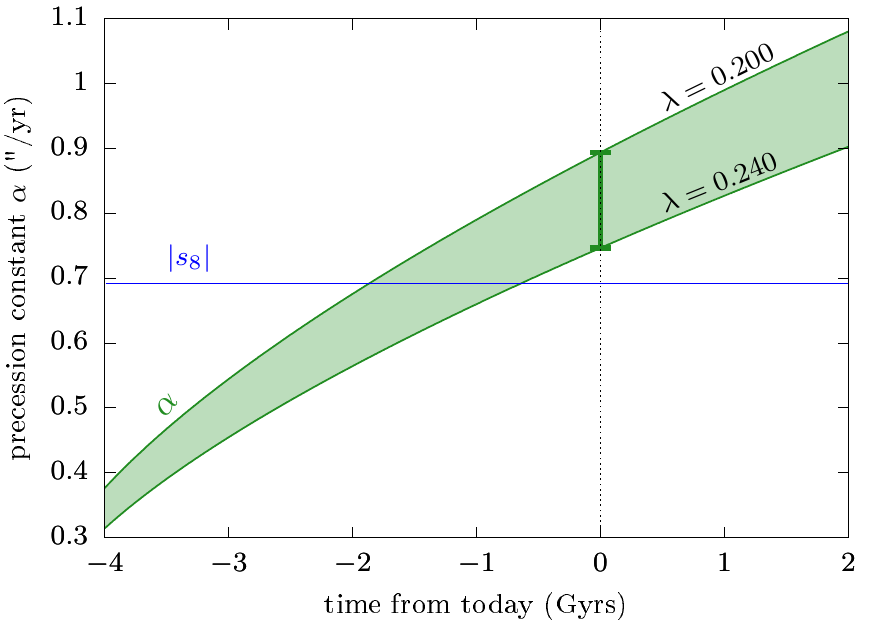}
      \caption{Evolution of the effective precession constant of Saturn due to the migration of Titan (adapted from \citealp{Saillenfest-etal_2020b}). The top and bottom green curves correspond to the two extreme values of the normalised polar moment of inertia $\lambda$ considered in this article. They appear into $\alpha$ through Eq.~\eqref{eq:alpha}. Both curves are obtained using the nominal value $b=1/3$ in Eq.~\eqref{eq:aTit}. Today's interval corresponds to the one shown in Fig.~\ref{fig:widths}; it is independent of the value of $b$ considered. The blue line shows Neptune's nodal precession mode, which was higher before the end of the late planetary migration.}
      \label{fig:alphaevol}
   \end{figure}

\begin{figure}
      \centering
      \includegraphics[width=\columnwidth]{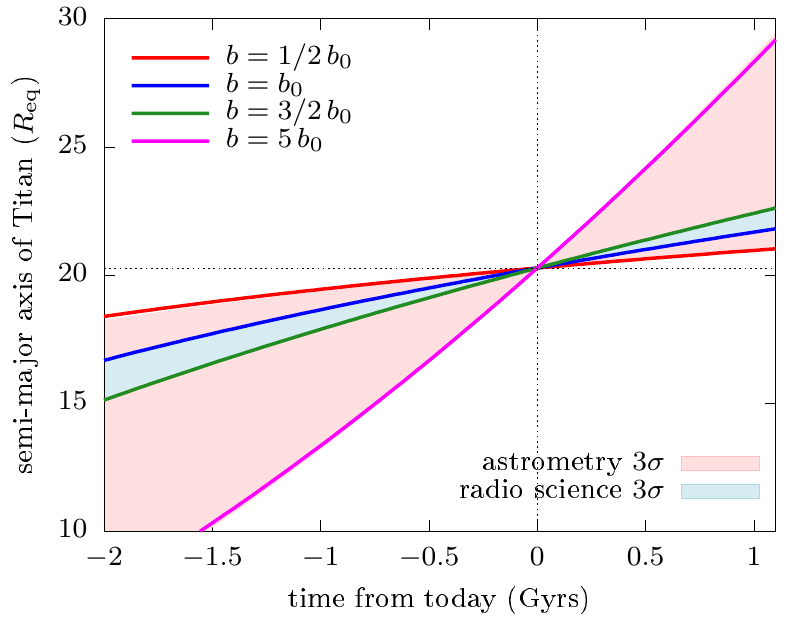}
      \caption{Time evolution of Titan's semi-major axis for different migration rates. The pink and blue intervals show the $3\sigma$ uncertainty ranges of astrometric and radio-science measurements, respectively \citep{Lainey-etal_2020}. The coloured curves are obtained by varying the parameter $b$ in Eq.~\eqref{eq:aTit}.}
      \label{fig:aTit}
   \end{figure}

   Because of tidal dissipation, satellites migrate over time. This produces a drift of the precession constant $\alpha$ on a timescale that is much larger than the precession motion (i.e. the circulation of $\psi$). The long-term spin-axis dynamics of a planet with migrating satellites is described by the Hamiltonian in Eq.~\eqref{eq:Hinit} where $\alpha$ is a slowly-varying function of time. Since Titan is in the close-in regime, its outward migration produces an increase in $\alpha$. The migration rate of Titan recently measured by \cite{Lainey-etal_2020} supports the tidal theory of \cite{Fuller-etal_2016}, through which the time evolution of Titan's semi-major axis can be expressed as
   \begin{equation}\label{eq:aTit}
      a_6(t) = a_0\left(\frac{t}{t_0}\right)^b\,,
   \end{equation}
   where $a_0$ is Titan's current mean semi-major axis, $t_0$ is Saturn's current age, and $b$ is a real parameter (see \citealp{Lainey-etal_2020}). Even though Eq.~\eqref{eq:aTit} only provides a crude model for Titan's migration, the parameter $b$ can be directly linked to the observed migration rate, independently of whether Eq.~\eqref{eq:aTit} is valid or not\footnote{In the latter case, $b$ should be considered as a non-constant quantity and what we measure today would only be its current value.}. Equation~\eqref{eq:aTit} implies that Titan's current tidal timescale $t_\mathrm{tide}=a_6/\dot{a}_6$ relates to $b$ as $b=t_0/t_\mathrm{tide}$. Considering a $3\sigma$ error interval, the astrometric measurements of \cite{Lainey-etal_2020} yield values of $b$ ranging in $[0.18,1.71]$, while their radio-science experiments yield values ranging in $[0.34,0.49]$. For the long-term evolution of Saturn's satellites, they adopt a nominal value of $b_0=1/3$, which roughly matches the observed migration of all satellites studied. Using this nominal value, we obtain a drift of the precession constant $\alpha$ as depicted in Fig.~\ref{fig:alphaevol}. Taking $b$ as parameter, a migration $n$ times faster for Titan is obtained by using in Eq.~\eqref{eq:aTit} a parameter $b=n\,b_0$. The corresponding evolution of Titan's semi-major axis is illustrated in Fig.~\ref{fig:aTit}.
   
   As mentioned by \cite{Saillenfest-etal_2020}, other parameters in Eq.~\eqref{eq:alpha} probably slightly vary over billions of years, such as the spin velocity of Saturn or its oblateness. We consider that the impact of their variations is small compared to the effect of Titan's migration (see Fig.~\ref{fig:alphaevol}) and contained within our exploration range. Moreover, all satellites, and not only Titan, migrate over time. However, being Titan so much more massive, its fast migration is by far the dominant cause of the drift of $\alpha$. Since its exact migration rate is still uncertain (see Fig.~\ref{fig:aTit}), this justifies our choice to only include Titan in Eq.~\eqref{eq:J2tilde}, while the use of its slightly increased mass $\tilde{m}_6$ yet allows us to obtain the right value of today's precession constant $\alpha$, as if all satellites were included.
   
   \subsection{Current spin orientation}\label{ssec:condinit}
   The initial orientation of Saturn's spin axis is taken from the solution of \cite{Archinal-etal_2018} averaged over short-period terms. With respect to Saturn's secular orbital solution (see Sect.~\ref{ssec:orbitsol}), this gives an obliquity $\varepsilon = 26.727^\circ$ and a precession angle $\psi=6.402^\circ$ at time J2000. The uncertainty on these values is extremely small compared to the range of $\alpha$ considered (see Sect.~\ref{ssec:alpha}).
   
\section{The past obliquity of Saturn: Exploration of the parameter space}\label{sec:backward}

    \subsection{Overview of possible trajectories}\label{ssec:over}
\begin{figure*}
      \includegraphics[width=\textwidth]{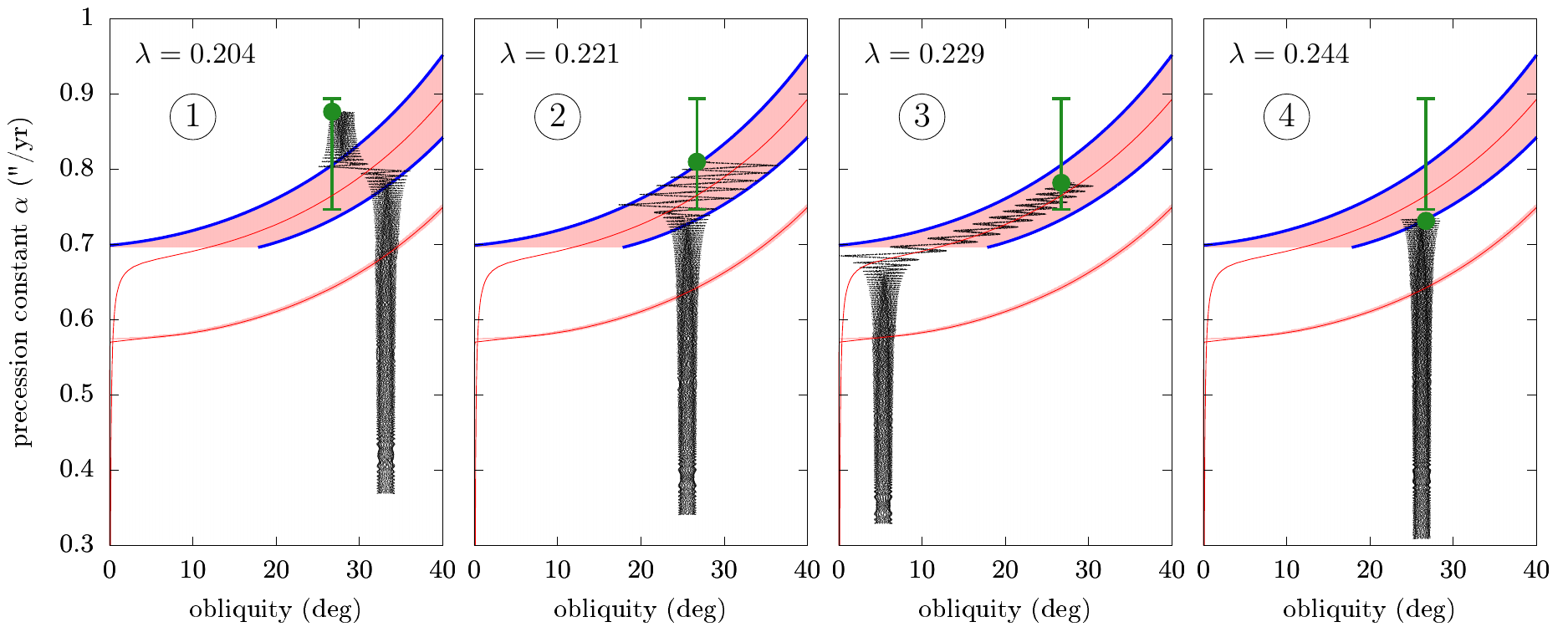}
      \caption{Examples illustrating the four different types of past obliquity evolution of Saturn. Each graph shows a $4$-Gyr numerical trajectory (black dots) computed for Titan's nominal migration rate and for a given value of Saturn's normalised polar moment of inertia $\lambda = C/(MR_\mathrm{eq}^2)$ specified in title. Today's location of Saturn is represented by the big green spot; the vertical error bar corresponds to our full exploration interval of $\lambda$. The red curves show the centre of first-order secular spin-orbit resonances (Cassini state 2) and the coloured areas represent their widths (same as Fig.~\ref{fig:widths}). The top large area is the resonance with $\phi_3$ and the bottom thin area is the resonance with $\phi_{19}$ (see Table~\ref{tab:zetashort}). The separatrices of the $\phi_3$ resonance are highlighted in blue. Going forwards in time, the trajectories go from bottom to top.}
   \label{fig:types}
   \end{figure*}

\begin{figure}
      \centering
      \includegraphics[width=\columnwidth]{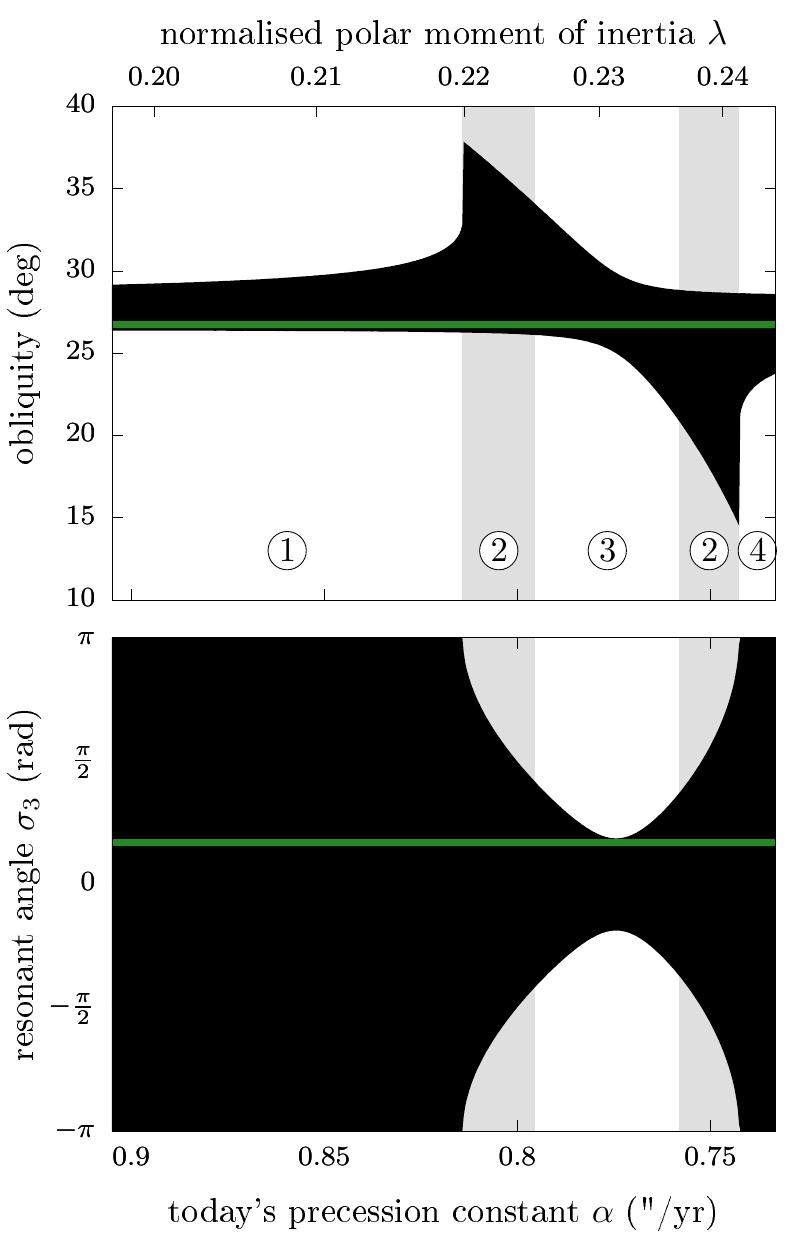}
      \caption{Current dynamics of Saturn's spin axis according to its normalised polar moment of inertia $\lambda$. The value of $\lambda$ (top horizontal axis) is linked to the current precession constant of Saturn (bottom horizontal axis) through Eqs.~\eqref{eq:alpha} and~\eqref{eq:J2prime}. The black interval shows the `instantaneous' oscillation range of Saturn's spin axis (i.e. without drift of $\alpha$) obtained by numerical integration. The resonant angle is $\sigma_3 = \psi+\phi_3$ (see Sect.~\ref{sec:dyn}). The green line shows Saturn's current obliquity and resonant angle. The background colour indicates the type of past evolution as labelled in the top panel (see text for the numbering).}
      \label{fig:currentstate}
   \end{figure}

   From the results of their backward numerical integrations, \cite{Saillenfest-etal_2020b} find that Saturn can have evolved through distinct kinds of evolution, which had previously been described by \cite{Ward-Hamilton_2004}. These different kinds of evolution are set by the outcomes of the resonant encounter between Saturn's spin-axis precession and the nodal precession mode of Neptune (term $\phi_3$ in Table~\ref{tab:zetashort} and largest resonance in Fig.~\ref{fig:widths}). The four possible types of past evolution are illustrated in Fig.~\ref{fig:types} for $b=b_0$. They are namely:
   \begin{itemize}
      \item Type~1: For $\lambda\leqslant 0.220$, Saturn went past the resonance through its hyperbolic point.
      \item Type~2: For $\lambda\in(0.220,0.224)\cup(0.237,0.241)$, Saturn was captured recently by crossing the separatrix of the resonance and followed the drift of its centre afterwards.
      \item Type~3: For $\lambda\in[0.224,237]$, the separatrix of the resonance appeared around Saturn's trajectory resulting in a $100\%$-sure capture at low obliquity. Saturn followed the drift of its centre afterwards.
      \item Type~4: For $\lambda\geqslant 0.241$, Saturn did not reach yet the resonance.
   \end{itemize}
   Figure~\ref{fig:currentstate} shows the current oscillation interval of Saturn's spin axis in all four cases. Trajectories of Type~3 are those featuring the smallest libration amplitude of the resonant angle~$\sigma_3$ and allowing for the smallest past obliquity of Saturn. Type~4 is ruled out by our uncertainty range for $\lambda$.
   
\begin{figure}
      \centering
      \includegraphics[width=\columnwidth]{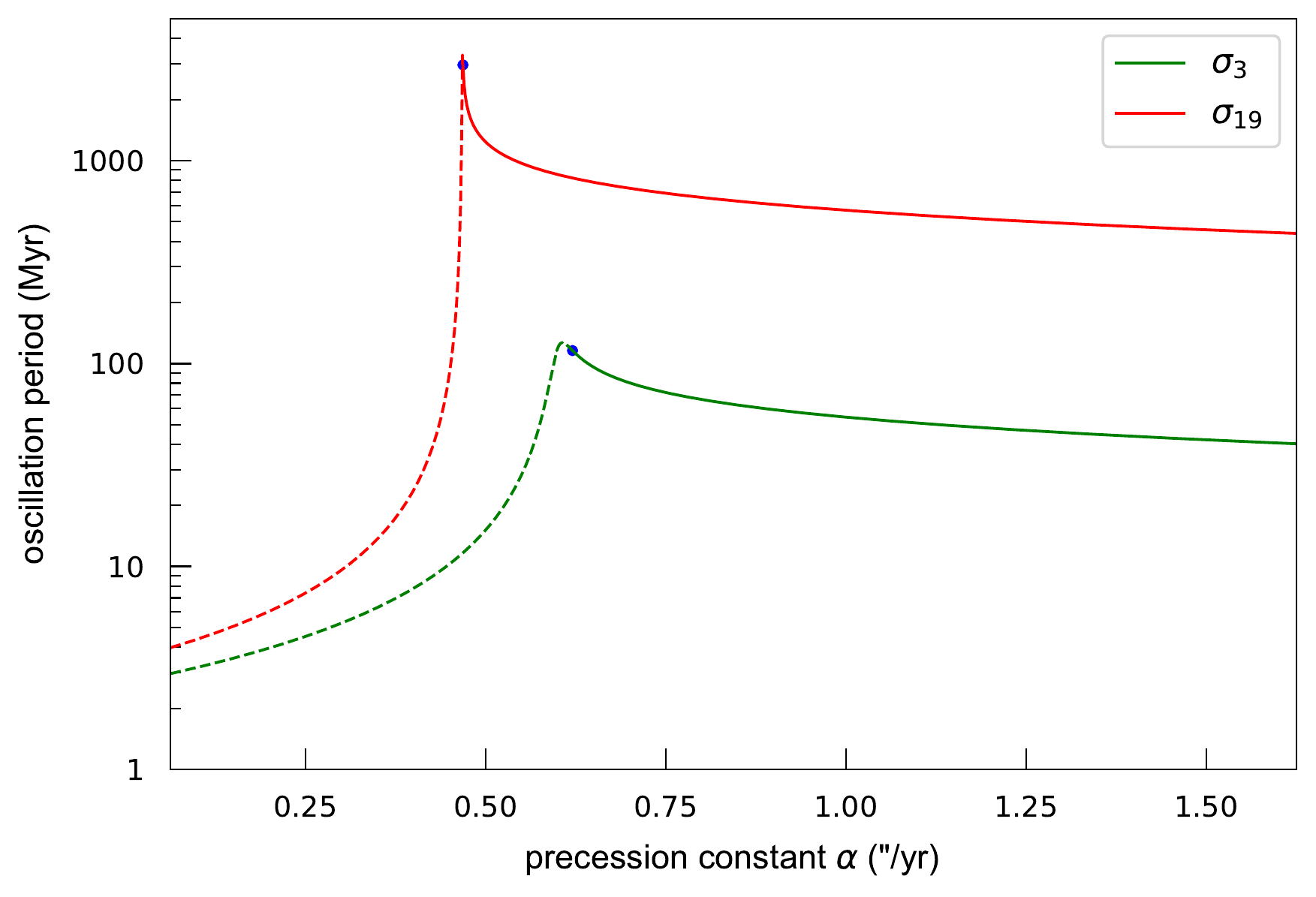}
      \caption{Period of small oscillations about the resonance centre for a resonance with $\phi_3$ or $\phi_{19}$. The resonant angles are $\sigma_3=\psi+\phi_3$ and $\sigma_{19}=\psi+\phi_{19}$, respectively. Dashed curves are used for oscillations about Cassini state 2 before the separatrix appears. The appearance of the separatrix is marked by a blue dot.}
      \label{fig:Plib}
   \end{figure}

   During its past evolution, Saturn also crossed a first-order secular spin-orbit resonance with the term $\phi_{19}$ which has frequency $g_7-g_8+s_7$ (see Table~\ref{tab:zetashort}). As shown in Fig.~\ref{fig:types}, however, this did not produce any noticeable change in obliquity for Saturn. Indeed, since this resonance is very small, the oscillation timescale of $\sigma_{19}=\psi+\phi_{19}$ inside the resonance is dramatically longer than the duration of the resonance crossing. This results in a short non-adiabatic crossing. The difference of oscillation timescales of $\sigma_3$ and $\sigma_{19}$ can be appreciated in Fig.~\ref{fig:Plib}. It explains why these two resonances have a so dissimilar influence on Saturn's spin-axis dynamics. This phenomenon has been further discussed by \cite{Saillenfest-etal_2020} in the case of Jupiter.
   
   \subsection{Adiabaticity of Titan's migration}\label{ssec:adiab}
   If the drift of $\alpha$ over time was perfectly adiabatic (i.e. infinitely slow compared to the oscillations of $\sigma_3$), the outcome of the dynamics would not depend on the exact migration rate of Titan; the latter would only affect the evolution timescale. In the vicinity of Titan's nominal migration rate, \cite{Saillenfest-etal_2020b} show that the drift of $\alpha$ is almost an adiabatic process. Here, we extend the analysis to a larger interval of migration rates in order to determine the limits of the adiabatic regime.
   
\begin{figure*}
      \centering
      \includegraphics[width=\textwidth]{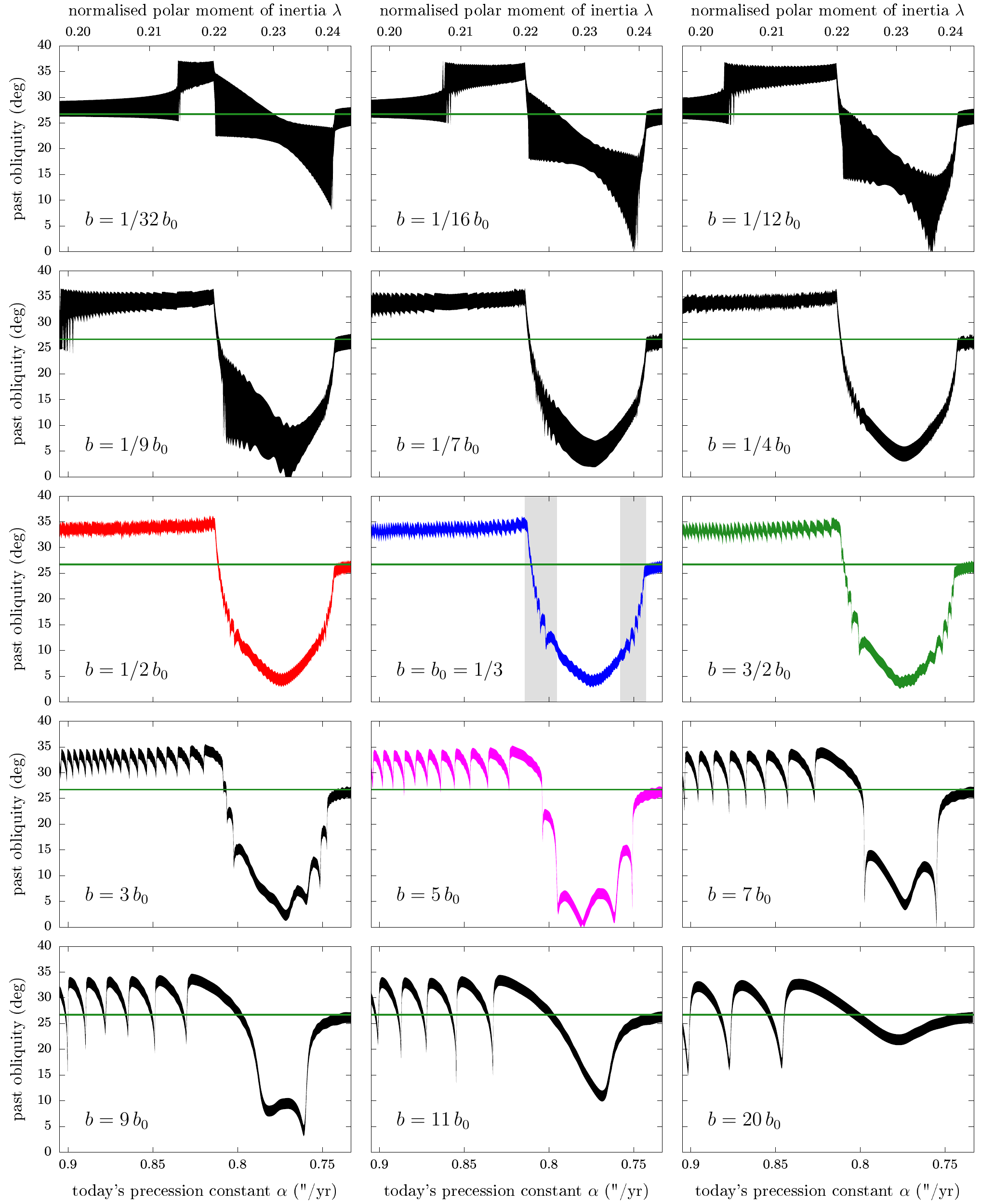}
      \caption{Past obliquity of Saturn for different migration rates of Titan. The top and bottom horizontal axes are the same as in Fig.~\ref{fig:currentstate} and the horizontal green line shows Saturn's current obliquity. For a given value of the normalised polar moment of inertia $\lambda$ (top horizontal axis), the curve width shows the oscillation range of obliquity $4$~Gyrs in the past obtained by backward numerical integration. The migration rates are labelled on each panel as a fraction of the nominal rate of \cite{Lainey-etal_2020}. The four coloured curves correspond to the migration rates illustrated in Fig.~\ref{fig:aTit}. The grey stripes in the central panel highlight trajectories of Type~2 (same as in Fig.~\ref{fig:currentstate}). The value of $b$ in the top left panel corresponds to a current quality factor $Q$ equal to $5000$ (see \citealp{Lainey-etal_2020}).}
      \label{fig:slowfast}
   \end{figure*}

   Figure~\ref{fig:slowfast} shows Saturn's obliquity $4$~Gyrs in the past obtained by backward numerical integrations for different migration rates of Titan and using values of $\lambda$ finely sampled in its exploration interval. Migration rates comprised between the red and magenta curves are compatible with the astrometric measurements of \cite{Lainey-etal_2020}, and migration rates comprised between the blue and green curves are compatible with their radio-science experiments (same colour code as in Fig.~\ref{fig:aTit}). As argued by \cite{Saillenfest-etal_2020b}, Titan's migration may have been sporadic, in which case $b$ would vary with time and the result would roughly correspond to a mix of several panels in Fig.~\eqref{fig:slowfast}. However, because of our current lack of knowledge about tidal dissipation mechanisms, refined evolution scenarios would only be speculative at this stage.
   
   The blue curve of Fig.~\ref{fig:slowfast} confirms that the nominal migration rate of \cite{Lainey-etal_2020} is close to the adiabatic regime, since smaller rates give very similar results (see the curves for a migration two times and four times slower). Non-adiabatic signatures are only substantial in the grey areas, that is, for recently captured trajectories that crossed the resonance separatrix (evolution Type~2). Indeed, the teeth-shaped structures are due to `phase effects', meaning that the precise outcome depends on the value of the resonant angle $\sigma_3$ during the separatrix crossing. For smaller migration rates, these structures pack together and tend to a smooth interval (that would be reached for perfect adiabaticity). If the migration of Titan is very slow, however, our $4$-Gyr backward integrations stop while Saturn is still close to the resonance, or even inside it. The curves obtained for $b\lesssim 1/7\,b_0$ have not enough time to completely relax from their initial shape shown in Fig.~\ref{fig:currentstate}. This means that if, as argued by \cite{Saillenfest-etal_2020b}, Titan is responsible for Saturn's current large obliquity, its migration cannot have been arbitrarily slow. Historical tidal models used to predict very small migration rates, as in the top left panel of Fig.~\ref{fig:slowfast}. Such small migration rates are unable to noticeably affect Saturn's obliquity over the age of the Solar System. This explains why previous studies considered that Saturn's precession constant remained approximatively constant since the late planetary migration \citep{Boue-etal_2009,Brasser-Lee_2015,Vokrouhlicky-Nesvorny_2015}. Figure~\ref{fig:slowfast} shows that for $\lambda\in[0.200,0.240]$, near-zero past obliquities can be achieved only if $b\gtrsim 1/16\,b_0$, that is, if Titan migrated by at least $1$~$R_\text{eq}$ after the late planetary migration. This condition is definitely achieved in the whole error ranges given by \cite{Lainey-etal_2020}, provided that Titan's migration did go on over a significant amount of time. Assuming that $b=b_0$, Titan should have migrated at least during several hundreds of million years before today in order for its semi-major axis to have changed by more than $1$~$R_\text{eq}$. On the contrary, no substantial obliquity variation could be produced if Titan only began migrating very recently (less than a few hundreds of million years) and always remained unmoved before that. As mentioned by \cite{Saillenfest-etal_2020b}, this extreme possibility appears unlikely but cannot be ruled out yet.
   
   When we increase Titan's migration rate above its nominal value, Fig.~\ref{fig:slowfast} shows that the adiabatic nature of the drift of $\alpha$ is gradually destroyed. For $b=3b_0$, phase effects become very strong and distort the whole picture. The magenta curve (which marks the limit of the $3\sigma$ error bar of \citealp{Lainey-etal_2020}) shows that the non-adiabaticity allows for a past obliquity of Saturn equal to exactly $0^\circ$. Such a null value is obtained when the oscillation phase of $\sigma_3$ brings Saturn's obliquity to zero exactly together with Cassini state 2. This configuration can only happen for finely tuned values of the parameters, which is why putting a primordial obliquity $\varepsilon\approx 0^\circ$ as a prerequisite puts so strong constraints on the parameter range allowed \citep{Brasser-Lee_2015,Vokrouhlicky-Nesvorny_2015}.
   
   If the resonance crossing is too fast, however, the resonant angle $\sigma_3$ has not enough time to oscillate before escaping the resonance. As a result, Saturn's spin-axis can only follow the drift of the resonance centre during a very limited amount of time, and only a moderate obliquity kick is possible. As discussed in Sect.~\ref{ssec:over}, this is what happens for the thin resonance with $\phi_{19}$. In Fig.~\ref{fig:slowfast}, the effect of overly fast crossings is clearly visible for $b\gtrsim 11b_0$. Beyond this approximate limit, all trajectories in our backward integrations cross the resonance separatrix, which means that trajectories of Type~3 are impossible and no small past obliquity can be obtained.
   
\begin{figure*}
      \centering
      \includegraphics[width=\textwidth]{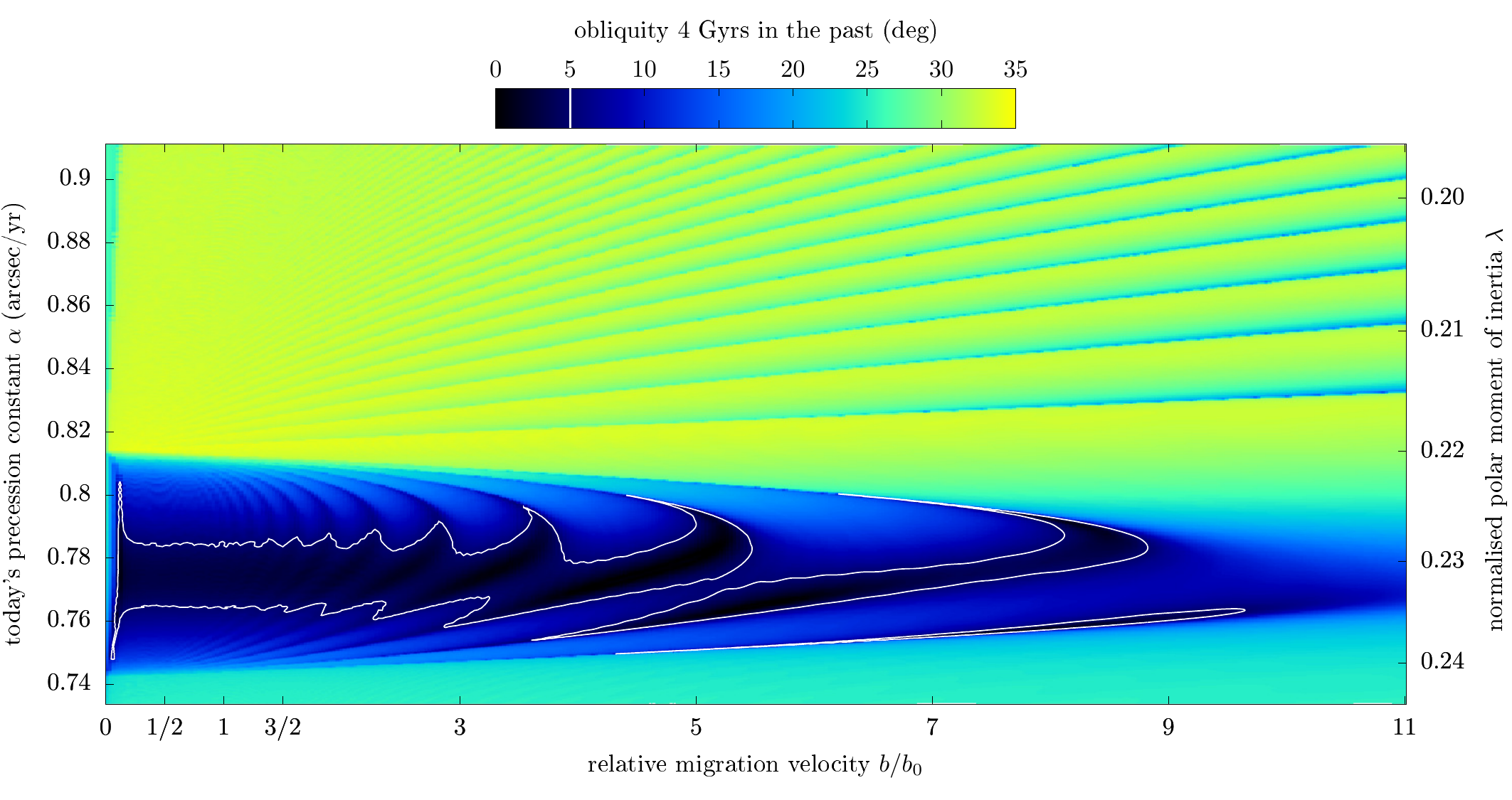}
      \caption{Past obliquity of Saturn as a function of Titan's migration velocity and Saturn's polar moment of inertia. Each panel of Fig.~\ref{fig:slowfast} corresponds here to a vertical slice. The colour scale depicts the minimum obliquity of the oscillation range, and the white curve highlights the $5^\circ$ level. The $3\sigma$ uncertainty ranges of \cite{Lainey-etal_2020} yield today approximately $b/b_0\in[1/2,5]$ for the astrometric measurements and $b/b_0\in[1,3/2]$ for the radio-science experiments (see Fig.~\ref{fig:aTit}).}
      \label{fig:obmin}
   \end{figure*}

   Figure~\ref{fig:obmin} summarises all values of Saturn's past obliquity obtained in our backward integrations as a function of Titan's migration rate and Saturn's polar moment of inertia. Non-adiabaticity is revealed by the coloured waves, denoting phase effects. As expected, the waves disappear for $b\lesssim b_0$: this is the adiabatic regime (see Fig.~4 of \citealp{Saillenfest-etal_2020b} for a zoom-in view). For very small migration rates, however, Titan would not have time in $4$~Gyrs to migrate enough to produce substantial effects on Saturn's obliquity. This is why the dark-blue region in Fig.~\ref{fig:obmin} does not reach $b=0$. For too fast migration rates, on the contrary, the resonance crossing is so brief that it can only produce a small obliquity kick. In particular no past obliquity smaller than $5^\circ$ is obtained for $b\gtrsim 10\,b_0$. This migration rate can therefore be considered as the largest one allowing Titan to be held responsible for Saturn's current large obliquity.
   
   \subsection{Extreme phase effects}\label{ssec:extreme}
   
   As can be guessed from the thinness of the spikes visible in some panels of Fig.~\ref{fig:slowfast}, the variety of outcomes obtained for trajectories that cross the resonance separatrix (i.e. Types~1 and 2) depend on the resolution used for sampling the parameter $\lambda$. The deepest spikes denote the strongest phase effects; they correspond to trajectories that reach the resonance almost exactly at its hyperbolic equilibrium point (called Cassini state~4: see e.g. \citealp{Saillenfest-etal_2019a} for phase portraits\footnote{There is a typographical error in \cite{Saillenfest-etal_2019a}: the list of the Cassini states given before Eq.~(22) should read (4,2,3,1) instead of (1,2,3,4) in order to match the denomination introduced by \cite{Peale_1969}.}). Since the resonance island slowly drifts as $\alpha$ varies over time, extreme phase effects can be produced when the hyperbolic point drifts away just as the trajectory gets closer to it, maintaining the trajectory on the edge between capture and non-capture into resonance. This kind of borderline trajectory is more common for strongly non-adiabatic drifts (i.e. the spikes in Fig.~\ref{fig:slowfast} are wider for larger $b$), because a faster drift of the resonance means that trajectories need to follow less accurately the separatrix in order to `chase' the hyperbolic point at the same pace as it gets away. If the drift of the resonance is too fast, however, trajectories are outrun by the resonance and strong phase effects are impossible. This is visible in the last panel of Fig.~\ref{fig:slowfast} (for $b=20\,b_0$), in which the spikes are noticeably smoothed.
   
\begin{figure}
      \centering
      \includegraphics[width=\columnwidth]{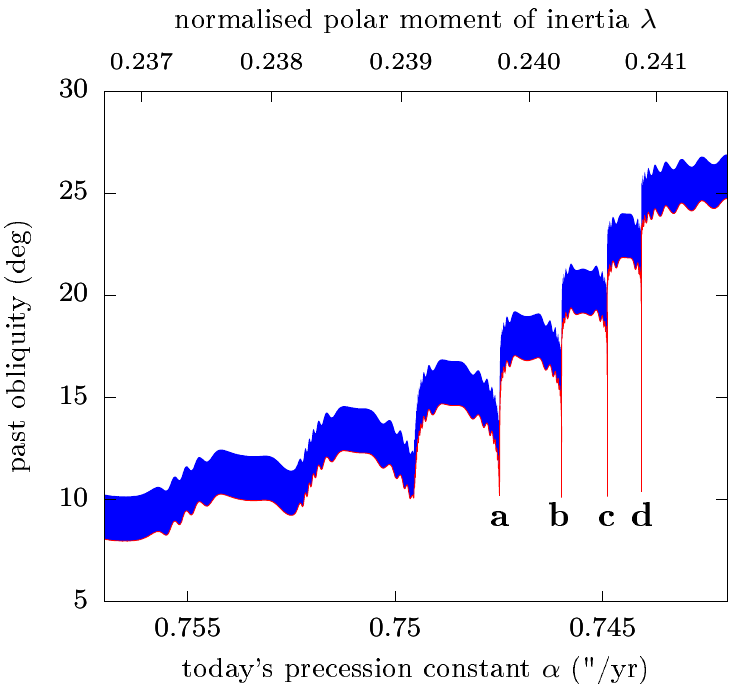}
      \caption{Zoom-in view of the central panel of Fig.~\ref{fig:slowfast}. We use a red curve to highlight the bottom limit of the blue interval, otherwise the narrowness of the spikes makes them invisible (the width of spike d is $\Delta\lambda\approx 10^{-14}$). This graph can be compared to Fig.~3 of \cite{Saillenfest-etal_2020b}, where such level of fine tuning is not shown due to its questionable physical relevance. See Appendix~\ref{asec:extreme} for examples of trajectories.}
      \label{fig:phase_effects}
   \end{figure}

   In order to investigate the outcomes of extreme phase effects, one can look for the exact tip of the spikes in Fig.~\ref{fig:slowfast} by a fine tuning of $\lambda$. For $b=b_0$ (central panel), a tuning of $\lambda$ at the $10^{-15}$ level shows that Type~2 trajectories all feature a minimum past obliquity of about $10^\circ$, as illustrated in Fig.~\ref{fig:phase_effects}. This minimum value is the same for each spike, and zooming in in Fig.~\ref{fig:phase_effects} shows that we do reach the bottom of the spikes. For Type~1 trajectories (i.e. $\lambda<0.220$ in the central panel of Fig.~\ref{fig:slowfast}), we managed to find past obliquities of about $28^\circ$ at the tip of the spikes, but using extended precision arithmetic may allow one to obtain even smaller values (possibly down to $10^\circ$ as for Type~2 trajectories). The width of these spikes ($\Delta\lambda<10^{-15}$) would however make them absolutely invisible in Figs.~\ref{fig:slowfast} and~\ref{fig:obmin}. In fact, the level of fine tuning required here is so extreme that such trajectories are unlikely to have any physical relevance. They are yet possible solutions in a mathematical point of view. Some examples are given in Appendix~\ref{asec:extreme}.
   
   These findings can be compared to previous studies, even though previous studies relied on a different tilting scenario. For a non-adiabatic drift of the resonance and a past obliquity fixed to $1.5^\circ$, \cite{Boue-etal_2009} found that if Saturn is not currently inside the resonance (i.e. if $\lambda<0.220$), an extremely narrow but non-zero range of initial conditions is able to reproduce Saturn's current orientation, with a probability less than $3\times10^{-8}$. Using a smaller set of simulations, \cite{Vokrouhlicky-Nesvorny_2015} did not even find a single of these trajectories. In light of our results, we argue that these unlikely trajectories are produced through the `extreme phase effects' described here. The vanishingly small probability of producing such trajectories is confirmed in Sect.~\ref{sec:CIsearch}.
   
\section{Monte Carlo search for initial conditions}\label{sec:CIsearch}
   
   In Sect.~\ref{sec:backward}, the past behaviour of Saturn's spin axis has been investigated using backward numerical integrations. If we now consider the space of all possible orientations for Saturn's primordial spin axis, each dynamical pathway has a given probability of being followed. A large subset of trajectories (those of Types~1 and~2) go through the separatrix of the large resonance with $\phi_3$. Separatrix crossings are known to be chaotic events (see e.g. \citealp{Wisdom_1985}), and since Saturn's orbital evolution is not restricted to its 3rd harmonic, the separatrix itself appears as a thin chaotic belt (see e.g. \citealp{Saillenfest-etal_2020}). Therefore, we can wonder whether the chaotic divergence of trajectories during separatrix crossings could lead to some kind of time-irreversibility in our numerical solutions (see e.g. \citealp{Morbidelli-etal_2020}), especially in the non-adiabatic regime, which has not been studied by \cite{Saillenfest-etal_2020b}. These aspects can be investigated through a Monte Carlo search for the initial conditions of Saturn's spin axis.
   
   \subsection{Capture probability}
   
\begin{figure*}
      \centering
      \includegraphics[width=\textwidth]{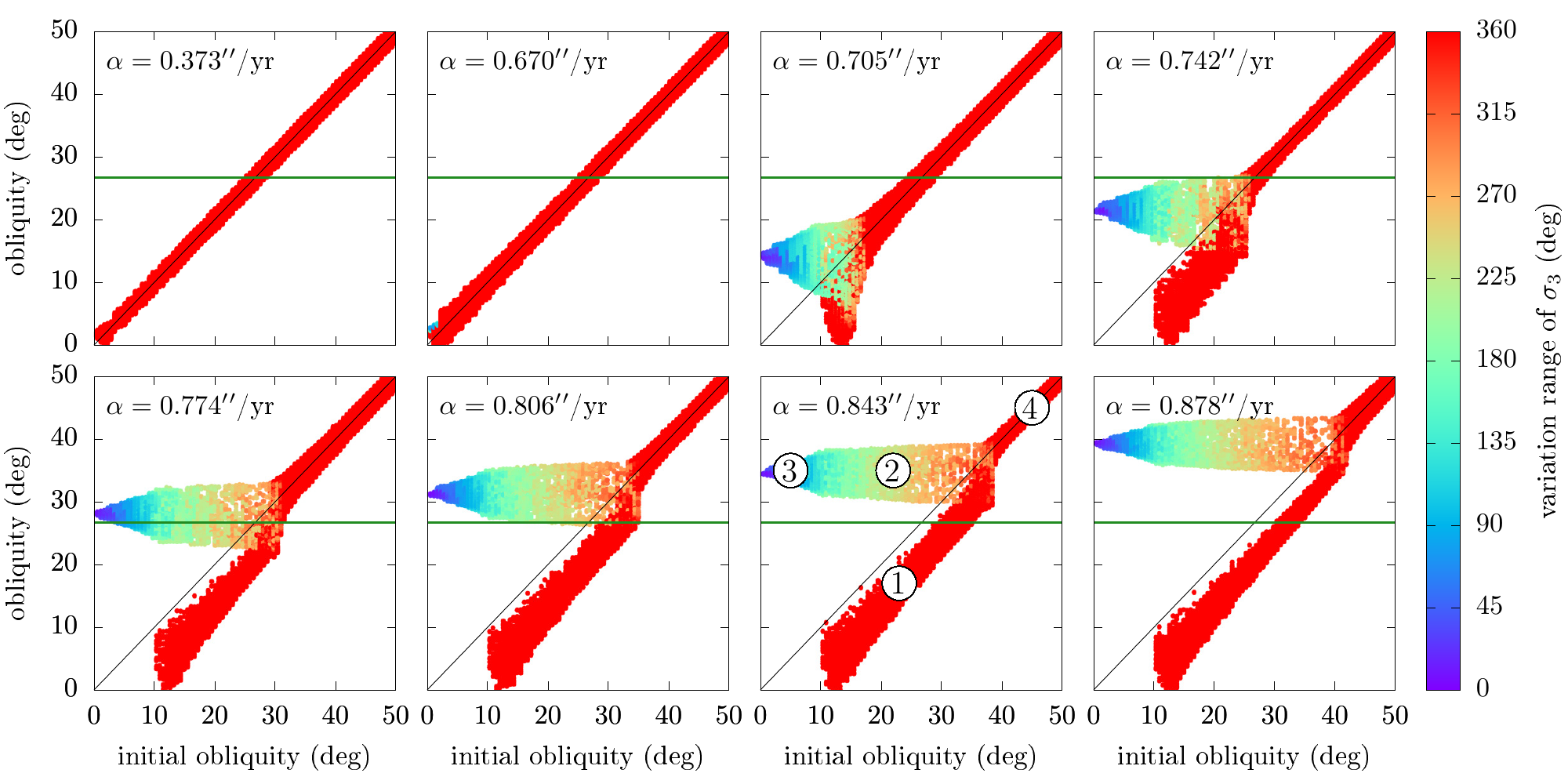}
      \caption{Snapshots of a Monte Carlo experiment computed for $\lambda=0.204$ and Titan's nominal migration rate. This experiment features $101$ values of initial obliquity between $0^\circ$ and $60^\circ$, for which $240$ values of initial precession angle are regularly sampled in $[0,2\pi)$. The value of $\alpha$ reached by the trajectories at the time of the snapshot is labelled on each panel. Each trajectory is represented by a small dot which is coloured according to the variation range of the resonant angle $\sigma_3$ (obtained by a $0.5$-Gyr numerical integration with constant $\alpha$). The horizontal green line shows the current obliquity of Saturn. At the beginning of the propagations, all trajectories are coloured red (since $\sigma_3$ circulates), and distributed along a diagonal line. Then, as $\alpha$ increases over time, trajectories are dispersed off the diagonal according to the four types of trajectories depicted in Fig.~\ref{fig:types} and labelled in the penultimate panel.}
      \label{fig:snap}
   \end{figure*}

   Our first experiment is designed as follows: for a given set of parameters $(b,\lambda)$, values of initial obliquity are regularly sampled between $0^\circ$ and $60^\circ$. Then, for each of those, we regularly sample values of initial precession angle $\psi\in[0,2\pi)$, and all trajectories are propagated forwards in time starting at $4$~Gyrs in the past (i.e. after the late planetary migration) up to today's epoch. Figure~\ref{fig:snap} shows snapshots of this experiment for $\lambda=0.204$ and Titan's nominal migration rate ($b=b_0$). The first snapshot is taken about $20$ million years after the start of the integrations, and the last snapshot is taken at today's epoch. Changing the value of $\lambda$ produces a shift of Saturn's precession constant $\alpha$ but no strong variation in its drift rate (see Fig.~\ref{fig:alphaevol}). Moreover, since this drift is almost an adiabatic process for $b=b_0$ (see Sect.~\ref{ssec:adiab}), a small change of drift rate does not modify the statistical outcome of the dynamics but only its timescale. For these reasons, a snapshot in Fig.~\ref{fig:snap} taken at a given time $t$ for $\lambda=0.204$ is undistinguishable from a snapshot taken at a slightly different time $\tilde{t}$ for another value $\tilde{\lambda}$. More precisely, if we introduce a function of time $f_\lambda(t)$ such that $t \longrightarrow \alpha = f_\lambda(t)$, an indistinguishable snapshot is obtained for a polar moment of inertia $\tilde{\lambda}$ at a time $\tilde{t} = f_{\tilde{\lambda}}^{-1}(\alpha)$. Hence, the only parameter that matters here is the value of the precession constant $\alpha$ reached by the trajectories. This is why the panels of Fig.~\ref{fig:snap} are labelled by $\alpha$ instead of $t$: this way they are valid for any value of $\lambda$.
   
   Before reaching the neighbourhood of the resonance with $\phi_3$, Fig.~\ref{fig:snap} shows that all trajectories only slightly oscillate around their initial obliquity value (compare the first two snapshots, taken for two very different values of $\alpha$). Then, as $\alpha$ continues to increase, the trajectories are gradually divided between the four possible types of evolution listed in Sect.~\ref{ssec:over}. All trajectories with initial obliquity smaller than about $10^\circ$ are captured in the resonance and lifted to high obliquities (Type~3: blue dots). Trajectories with a larger initial obliquity can either be captured (Type~2: green and orange dots) or go past the resonance through its hyperbolic point (Type~1: lowermost red dots).
   
\begin{figure}
	\centering
	\includegraphics[width=\columnwidth]{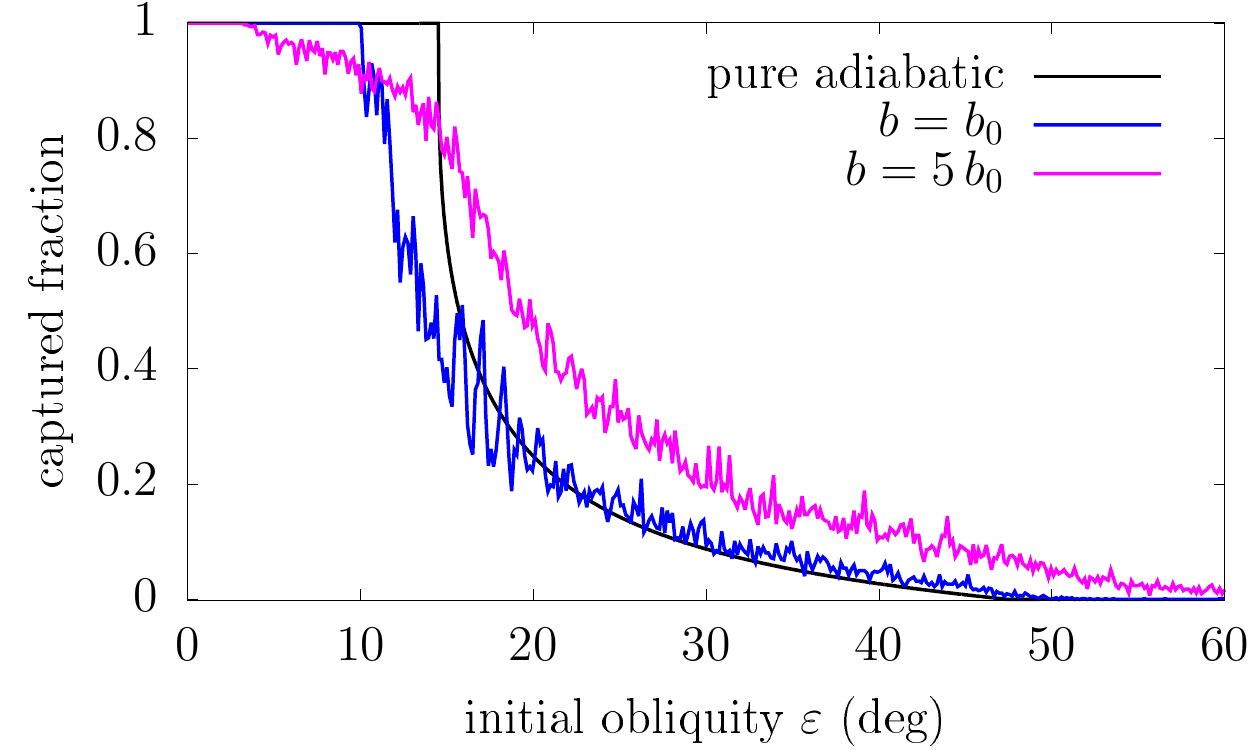}
	\caption{Capture probability of Saturn in secular spin-orbit resonance with $\phi_3$ as a function of its primordial obliquity. For each initial obliquity ($401$ values between $0$ and $60^\circ$), $720$ values of initial precession angle are uniformly sampled in $[0,2\pi)$ and propagated forwards in time starting from $-4$~Gyrs and until every trajectory has reached the resonance. This experiment is repeated with two different migration laws for Titan (see labels).  The result is virtually independent of the value chosen for $\lambda$. The fraction of captured trajectories (coloured curves) is compared to the perfect adiabatic case (black curve) computed with the analytical formulas of \cite{Henrard-Murigande_1987}.}
	\label{fig:proba}
\end{figure}

   Assuming that Saturn's primordial precession angle $\psi$ is a random number uniformly distributed in $[0,2\pi)$, the probability of capture in resonance is given by the fraction of points ending up in the pencil-shaped structure of Fig.~\ref{fig:snap}. The result is shown in Fig.~\ref{fig:proba}, in which we increased the resolution for better statistical significance. Assuming perfect adiabaticity, each outcome can be modelled as a probabilistic event ruled by analytical formulas (see \citealp{Henrard-Murigande_1987,Ward-Hamilton_2004,Su-Lai_2020}). As shown by Fig.~\ref{fig:proba}, non-adiabaticity tends to smooth the probability profile and to reduce the interval of $100\%$-sure capture. For growing initial obliquity, the probability of Type~2 trajectories (i.e. capture) decreases, favouring Type~1 trajectories instead (i.e. crossing without capture).
   
   \subsection{Loose success criteria: Probing all dynamical pathways}
   
\begin{figure}
      \centering
      \includegraphics[width=\columnwidth]{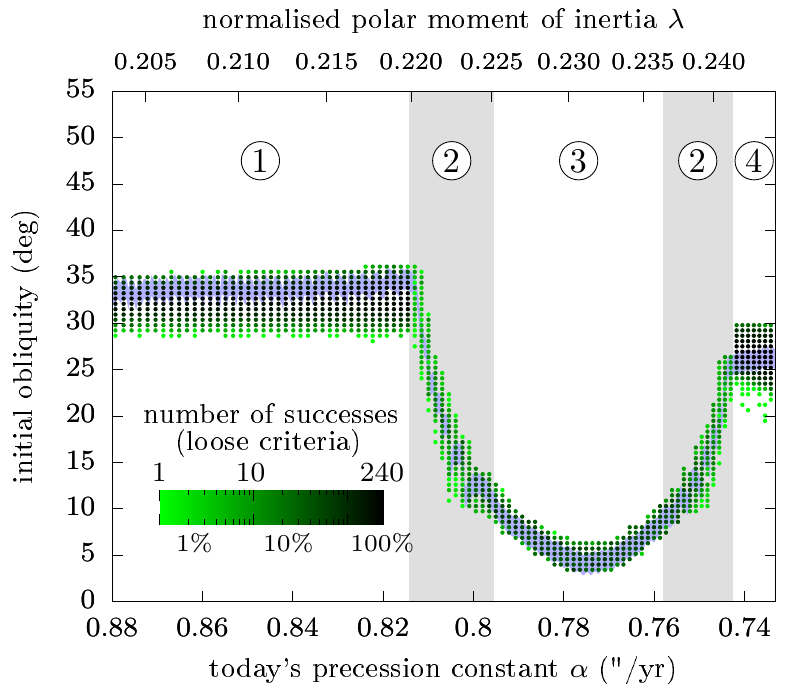}
      \caption{Brute-force search for Saturn's past obliquity using the loose success criteria: probing all dynamical pathways. We use Titan's nominal migration law ($b=b_0$). Among all trajectories evenly sampled in the space of the normalised polar moment of inertia $\lambda$ (top horizontal axis, $101$ values), of the initial obliquity (vertical axis, $101$ values), and of the initial precession angle ($240$ values between $0$ and $2\pi$), we only keep those matching Saturn's spin axis today according to our loose success criteria (see text). Each point is coloured according to the number of successful runs among the $240$ initial precession angles; the success ratio is written below the colour bar. A point is not drawn if no successful trajectory is found. In the back, the blue interval shows the past obliquity of Saturn obtained by backward numerical integration (same as Fig.~\ref{fig:slowfast} for $b=b_0$), showing the consistency between backward and forward integrations in time. The background stripes and their labels have the same meaning as in Fig.~\ref{fig:currentstate}.}
      \label{fig:CIsearch}
   \end{figure}

   Over all possible trajectories, we now look for those matching Saturn's actual spin-axis dynamics today. We first consider `loose success criteria', for which  a run is judged successful if: \emph{i)} Saturn's current obliquity $\varepsilon=26.727^\circ$ lies within the final spin-axis oscillation interval, and \emph{ii)} the libration amplitude of the resonant angle $\sigma_3$ lies within $5^\circ$ of the actual amplitude shown in Fig.~\ref{fig:currentstate}. These criteria are not chosen to be very strict in order to probe all dynamical pathways in the neighbourhood of Saturn's spin-axis orientation, including some that could have been missed by the backward propagations of Sect.~\ref{sec:backward}. Our results are depicted in Fig.~\ref{fig:CIsearch}. We closely retrieve the predictions of backward numerical integrations, in particular for trajectories of Type~3. Narrowing the target interval leads to an even better match. For trajectories of Type~2 (grey background), we obtain a larger spread of initial obliquities because our success criteria do not include any restriction on today's phase of the resonant angle $\sigma_3$, but only on its oscillation amplitude. The results shown in Fig.~\ref{fig:CIsearch} are therefore less shaped by `phase effects' discussed in Sect.~\ref{ssec:adiab}. Since a slight change in Titan's migration rate would result in a phase shift, we can interpret Fig.~\ref{fig:CIsearch} as encompassing different migration rates around the nominal observed rate. The results presented in Fig.~\ref{fig:CIsearch} are therefore more general than those obtained using backward numerical integrations. In accordance with Fig.~\ref{fig:proba}, the success ratio for Type~2 trajectories sharply decreases for increasing initial obliquity (colour gradient), because most initial conditions lead to a resonance crossing without capture. Moreover, we do not detect trajectories as extreme as those presented in Sect.~\ref{ssec:extreme} (i.e. with an initial obliquity of about $10^\circ$ all over the width of Zone~2), because they require initial conditions that are too specific for our sampling; the probability of obtaining one is indeed negligible. Finally, for trajectories of Types~1 and~4, which are today out of the resonance, our `loose success criteria' are extremely permissive, since the variation amplitude of $\sigma_3$ is $2\pi$ for all trajectories (see Fig.~\ref{fig:currentstate}). This explains why Fig.~\ref{fig:CIsearch} shows large intervals of black dots. These intervals can be spotted in Fig.~\ref{fig:snap}, where they appear as the whole range of red dots that are pierced by the green horizontal line.
   
   Figure~\ref{fig:CIsearch} does not feature unexpected dynamical paths that could have been missed by our backward integrations, even though signatures of chaos are visible in the sparse spreads of coloured dots. From this close match, we conclude that the chaos is not strong enough here to significantly mingle the trajectories and to produce a substantial phenomenon of numerical irreversibility. As one can point out, separatrix crossings would have been irreversible if, in order to predict the different outcomes, we used the adiabatic invariant theory instead of numerical integrations (see \citealp{Henrard-Murigande_1987,Ward-Hamilton_2004,Su-Lai_2020}). Indeed, in the adiabatic invariant theory, the resonant angle is assumed to oscillate infinitely faster than the drift of $\alpha$ and phase effects are modelled as probabilistic events \citep{Henrard_1982,Henrard_1993}. This probabilistic modelling of chaos explains why separatrix crossings are not reversible when using this theory.
   
   \subsection{Strict success criteria: Relative likelihood of producing Saturn's current state}\label{ssec:MCstrict}
   
\begin{figure}[h]
	  \centering
	  \includegraphics[width=\columnwidth]{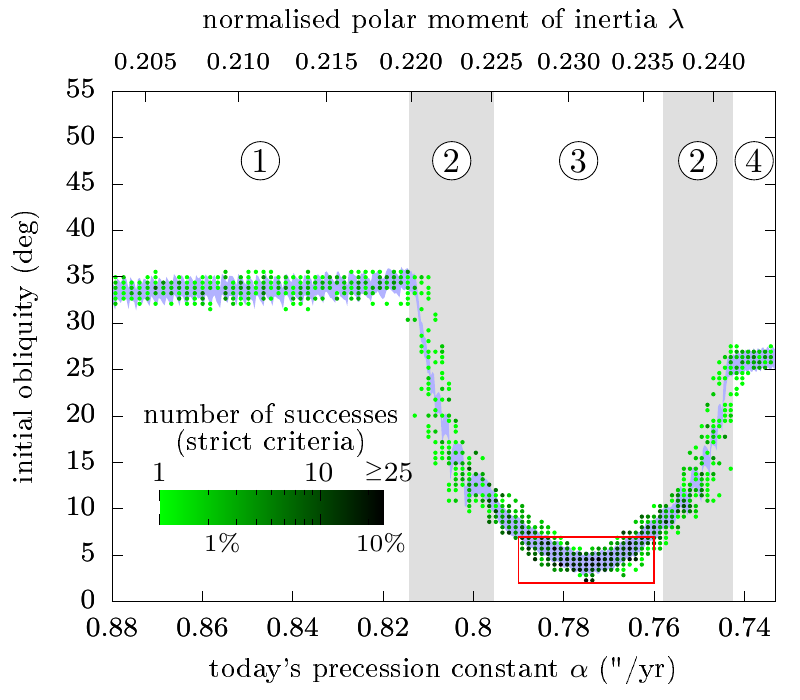}
	  \caption{Same as Fig.~\ref{fig:CIsearch}, but using the strict success criteria: comparing the relative likelihood of producing Saturn's current state. As in Fig.~\ref{fig:CIsearch}, each point of the graph is made of $240$ simulations with initial $\psi\in[0,2\pi)$. The red rectangle highlights the region featuring the highest success ratios.}
	  \label{fig:CIsearch_strong}
   \end{figure}

   In order to compare the likelihood of producing Saturn's current state in the space of all possible initial conditions, our loose success criteria are not enough. Independently of whether Saturn is inside or outside the resonance today, its spin-axis precession is not uniform, which means that the phase of Saturn's spin-axis motion at a given time is not uniformly distributed and must therefore be taken into account, too. Moreover, we saw in Sect.~\ref{ssec:adiab} that out of the strict adiabatic regime, phase effects (that are deliberately ignored by our loose success criteria) do matter to reproduce Saturn's current spin-axis orientation; actually, the very notion of `libration' loses its meaning when the drift of $\alpha$ is not adiabatic, since the resonance is distorted before $\sigma_3$ has time to perform a single cycle. For these reasons, we now define `strict success criteria', for which a run is judged successful if: \emph{i)} today's obliquity $\varepsilon$ lies within $0.5^\circ$ of the true value, and \emph{ii)} today's precession angle $\psi$ lies within $5^\circ$ of the true value. These criteria are very narrow, but still within reach of our millions of numerical propagations. The result is shown in Fig.~\ref{fig:CIsearch_strong} for Titan's nominal migration rate. As expected, the points are more sparse than in Fig.~\ref{fig:CIsearch} and the success ratios are smaller. Assuming that Saturn's primordial precession angle is a random number uniformly distributed between $0$ and $2\pi$, the colour gradient in Fig.~\ref{fig:CIsearch_strong} is a direct measure of the likelihood to reproduce Saturn's current state. Type~3 trajectories are greatly favoured: they feature the maximum likelihood, which is about ten times the likelihood of Type~1 trajectories. The region with maximum likelihood is for past obliquities between about $2^\circ$ and $7^\circ$, and current precession constant $\alpha$ between about $0.76$ and $0.79''\,$yr$^{-1}$ (red box).
   
   As already discussed by \cite{Ward-Hamilton_2004} and \cite{Hamilton-Ward_2004}, there are two reasons why Type~3 trajectories are the most likely: first, they have a $100\%$ chance of being captured inside the resonance (whereas Types~1 and 2 both have a non-zero probability of failure, see Fig.~\ref{fig:proba}); second, Type~3 trajectories oscillate today with a small amplitude inside the resonance, which means that all of them feature a similar value of the precession angle $\psi$, imposed by the resonance relation $\sigma_3\sim 0$. On the contrary, other types of trajectories either feature a large oscillation amplitude of $\sigma_3$ (Type~2) or circulation of $\sigma_3$ (Types~1 and 4); therefore, they only sweep over Saturn's actual orientation once in a while, and matching it today would only be a low-probability `coincidental' event\footnote{The same argument has been pointed out for Jupiter by \cite{Ward-Canup_2006} and \cite{Saillenfest-etal_2020}.}. As shown by Fig.~\ref{fig:CIsearch_strong}, the least favoured trajectories are those of Type~2, especially for high initial obliquities, because of the strong decrease in capture probability (see Fig.~\ref{fig:proba}).
   
   In order to explore all migration rates and bring further constraints on the model parameters, we now turn to a second Monte Carlo experiment, with the following approach: assuming that Saturn was indeed tilted as a result of Titan's migration, we look for the possible values of the parameters $(b,\lambda)$ allowed, with their respective likelihood. This approach is similar to those used in previous studies (e.g. \citealp{Vokrouhlicky-Nesvorny_2015}).
   
   The notion of likelihood associated with this second experiment deserves some comments. Since Saturn's spin axis performed many precession revolutions in $4$~Gyrs and since it was initially not locked in resonance, a tiny error in the model rapidly spreads over time into a uniform probability distribution of the precession angle $\psi$ in $[0,2\pi)$. This is the reason why, in absence of any mechanism able to maintain $\psi$ in a preferred direction, it is legitimate to consider a uniform initial distribution for $\psi$, as people usually do (and as we already did above). Establishing a prior distribution for $\varepsilon$, instead, is more hazardous: we know that near-zero values are expected from formation models, but small primordial excitations cannot be excluded. Such excitations could be attributed to the phase of planetesimal bombardment at the end of Saturn's formation or by abrupt resonance crossings stemming from the dissipation of
   the protoplanetary and/or circumplanetary discs (see e.g. \citealp{Millholland-Batygin_2019}). Therefore, we arbitrarily consider here values of initial obliquity $\varepsilon\lesssim 5^\circ$, which leaves room for a few degrees of primordial obliquity excitation. This choice is somewhat guided by the $3^\circ$-obliquity of Jupiter, a part of which could possibly be primordial \citep{Ward-Canup_2006,Vokrouhlicky-Nesvorny_2015}. Jupiter is located today near a secular spin-orbit resonance with $s_7$ (see Table~\ref{tab:zetashort}), but contrary to Saturn, its satellites did not migrate enough yet to substantially increase its obliquity \citep{Saillenfest-etal_2020}; however, in order to ascertain possible values for Jupiter's primordial obliquity, the effect of the past migration of the Galilean satellites would need to be studied. We choose to use a uniform random distribution of $\varepsilon$, resulting in a non-uniform distribution of spin-axis directions over the unit sphere that favours small obliquities\footnote{In order to uniformly sample the unit sphere, one should consider instead a uniform distribution of $\cos\varepsilon$.}. The influence of our arbitrary choice of Saturn's initial obliquity is discussed below.
   
   \begin{figure*}
      \centering
      \includegraphics[width=\textwidth]{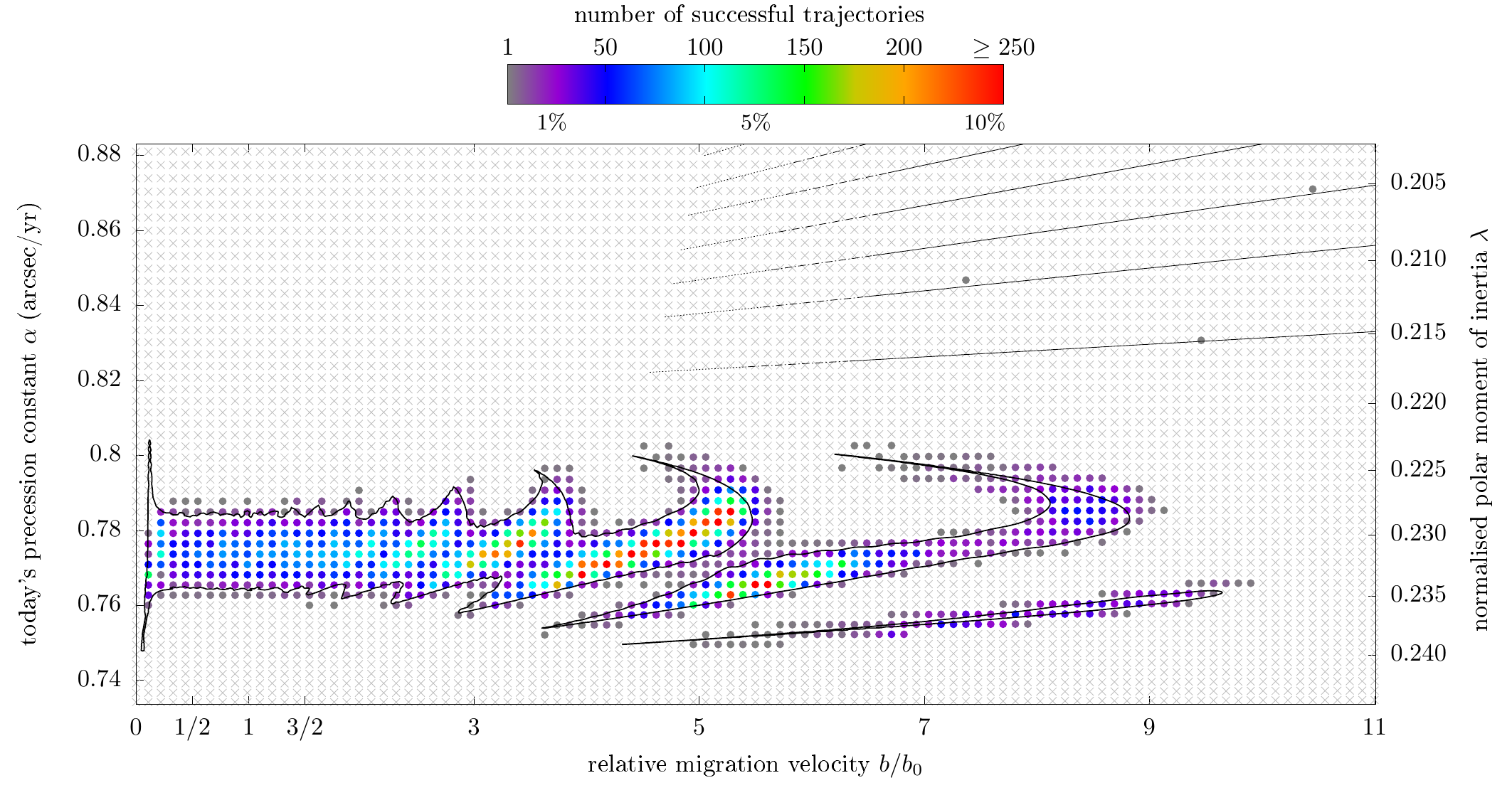}
      \caption{Distribution of the solutions starting from a low primordial obliquity and matching our strict success criteria. For each set $(b,\lambda)$ of the parameters, $2400$ values of initial obliquity $\varepsilon$ and precession angle $\psi$ are drawn from a uniform random distribution in $(\varepsilon,\psi)\in[0^\circ,5^\circ]\times[0,2\pi)$. Coloured dots show the parameter sets $(b,\lambda)$ for which at least one successful trajectory was found; the success ratio is written below the colour bar. Light-grey crosses mean that no successful trajectory was found over our $2400$ initial conditions. The black contours show the $5^\circ$-level obtained through backward numerical integrations (same as Fig.~\ref{fig:obmin}), showing the consistency between backward and forward integrations in time. The black lines in the top portion show the approximate location of the border of the blue stripes in Fig.~\ref{fig:obmin}, where extreme phase effects can happen; the corresponding ranges of parameters are so narrow that they are missed by the resolution of Fig.~\ref{fig:obmin} (see Sect.~\ref{ssec:extreme}).}
      \label{fig:constraint}
   \end{figure*}
   
   In practice, our setup is the following: over a grid of point $(b,\lambda)$ of the parameter space, we perform each time $2400$ numerical integrations starting from random initial conditions $(\varepsilon,\psi)$ with $\varepsilon\leqslant 5^\circ$ and $\psi\in[0,2\pi)$. All trajectories are then propagated from $-4$~Gyrs up to today's epoch, and we only keep trajectories matching Saturn's current spin-axis orientation according to our strict success criteria. Figure~\ref{fig:constraint} shows the result of this experiment. Again, we closely retrieve the predictions of backward integrations from Sect.~\ref{sec:backward}, confirming the reversible nature of the dynamics, and helping us to interpret the patterns obtained. The wavy structure at $3b_0\lesssim b\lesssim 5b_0$ resembles to some extent the successful matches of \cite{Vokrouhlicky-Nesvorny_2015}, reminding us that the basic dynamical ingredients are the same, even though the mechanism producing the resonance encounter in their study is different (their Fig.~7 is rotated clockwise). Unsurprisingly, the highest concentrations of matching trajectories in Fig.~\ref{fig:constraint} are located in the regions where backward propagations result in near-zero primordial obliquities (compare with Fig.~\ref{fig:obmin}). The maximum likelihood thus favours slightly non-adiabatic migration rates, for $b$ lying roughly between $3b_0$ and $6b_0$. According to \cite{Lainey-etal_2020}, such values are consistent with the $3\sigma$ uncertainty ranges of Titan's current migration rate obtained from astrometric measurements ($b/b_0\in[1/2,5]$), but not with the uncertainty ranges given by radio-science experiments ($b/b_0\in[1,3/2]$). However, successful trajectories with substantial likelihood are anyway found in a very large interval of migration rates, which extends much farther than the uncertainty range of \cite{Lainey-etal_2020}. We can therefore not bring any decisive constraint on Titan's migration history that would be tighter than those obtained from observations. Yet, the tilting of Saturn would impose strong constraint on Saturn's polar moment of inertia. As already visible in the figures of \cite{Saillenfest-etal_2020b}, tilting Saturn from $\varepsilon\lesssim 5^\circ$ in the adiabatic regime would require that $\lambda$ lies between about $0.228$ and $0.235$. Figure~\ref{fig:constraint} shows that allowing for non-adiabatic effects ($b\gtrsim 3b_0$) widens this range to about $[0.224,0.239]$.
   
   Interestingly, Fig.~\ref{fig:constraint} features three trajectories affected by an `extreme phase effect' (see Sect.~\ref{ssec:extreme}), visible as the three isolated grey points at $\lambda\approx 0.205$, $0.210$, and $0.215$. These trajectories are of Type~1 (i.e. currently out of the resonance) and fit our strict success criteria. The existence of these points recalls that such trajectories are extremely rare (we found only three over millions of trials), but yet possible, as previously reported by \cite{Boue-etal_2009}. They correspond to the narrow dark edges of the blue stripes in the top portion of Fig.~\ref{fig:obmin}. The complete trajectory producing the leftmost of these points can be found in Appendix~\ref{asec:extreme}.
   
\begin{figure*}
      \centering
      \includegraphics[width=\textwidth]{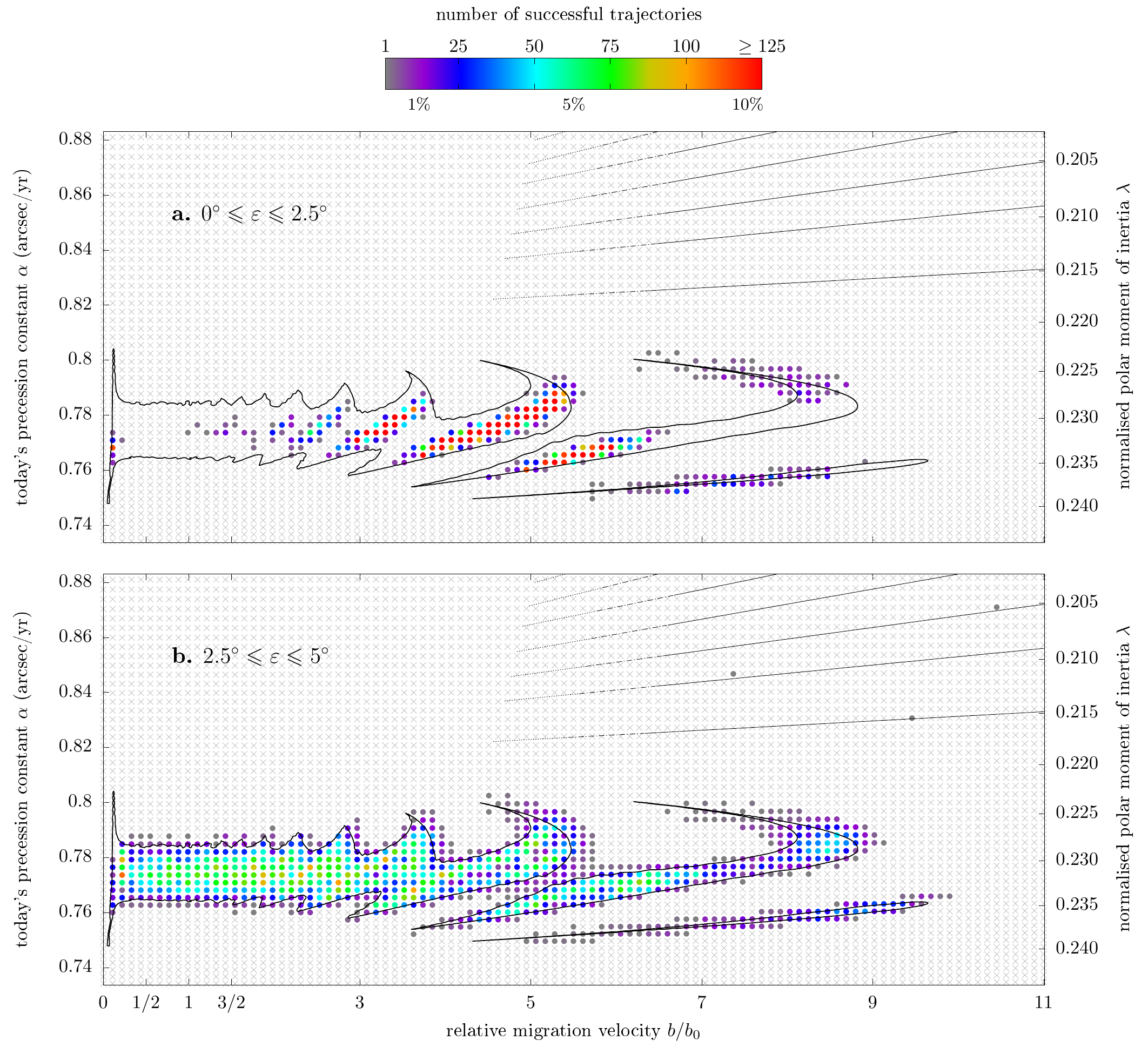}
      \caption{Same as Fig.~\ref{fig:constraint}, but for statistics based on a sub-sample of simulations. \textbf{a}: initial conditions in $(\varepsilon,\psi)\in[0^\circ,2.5^\circ]\times[0,2\pi)$. \textbf{b}: initial conditions in $(\varepsilon,\psi)\in[2.5^\circ,5^\circ]\times[0,2\pi)$. In both panels, each point is made of about $1200$ initial conditions extracted from the simulations from Fig.~\ref{fig:constraint}.}
      \label{fig:constraint-halves}
   \end{figure*}

   As mentioned above, the likelihood measure depicted in Fig.~\ref{fig:constraint} is conditioned by our assumptions about the initial value of $\varepsilon$. The influence of these assumptions can be investigated from our large simulation set. Figure~\ref{fig:constraint-halves} shows the statistics restricted to the lower- and higher-obliquity halves of the distribution. Restricting the initial obliquity to $\varepsilon\lesssim 2.5^\circ$ suppresses most successful matches from the adiabatic regime ($b\lesssim 3b_0$), as one could have guessed from previous figures. On the contrary, restricting the statistics to the upper half of the distribution ($\varepsilon\in[2.5^\circ,5^\circ]$) greatly shifts the point of maximum likelihood towards the adiabatic regime. Further experiments are provided in Appendix~\ref{ssec:prior} with initial obliquity values up to $10^\circ$. These experiments show that the adiabatic and non-adiabatic regimes are roughly equally likely if one considers an isotropic distribution of initial spin orientations (with $\varepsilon\lesssim 5^\circ$ or $\varepsilon\lesssim 10^\circ$) instead of a distribution favouring small initial obliquities as in Fig.~\ref{fig:constraint}. Unsurprisingly, the adiabatic regime and Titan's nominal migration rate are the most likely if one considers initial obliquity values as $2^\circ\lesssim\varepsilon\lesssim 7^\circ$ (i.e. in the red box of Fig.~\ref{fig:CIsearch_strong}).
   
   This discussion shows how important is the prior chosen for the initial conditions. Assumption biases are unavoidable and were also present in previous studies: \cite{Boue-etal_2009} assumed $\varepsilon=1.5^\circ$; \cite{Vokrouhlicky-Nesvorny_2015} assumed $\varepsilon=0.1^\circ$ (with respect to the orbit averaged over all angles but $\phi_3$); and \cite{Brasser-Lee_2015} assumed  $\varepsilon\approx 0.05^\circ$ (with respect to the invariable plane, i.e. the orbit averaged over all $\phi_k$). As shown by our results, leaving room for a few degrees of extra primordial excitation, or even only $0.5^\circ$, in any of those studies could have greatly enhanced the chances of success. As recalled above, a few degrees of primordial obliquity excitation are plausible and could be explained in different ways. In this regard, the most general overview of our findings is given by Fig.~\ref{fig:obmin}, since it does not presuppose any initial obliquity for Saturn, and Fig.~\ref{fig:CIsearch_strong} shows the respective likelihood of each dynamical pathway, still with no assumption about the initial obliquity.
   
\section{The future obliquity of Saturn}\label{sec:fut}
   Since Titan goes on migrating today, Saturn's obliquity is likely to continuously vary over time. Hence, we can wonder whether it could reach large values, in the same way as Jupiter \citep{Saillenfest-etal_2020}. In order to explore the future obliquity dynamics of Saturn, we propagate Saturn's spin-axis from today up to $5$~Gyrs in the future.
   
\begin{figure}
      \centering
      \includegraphics[width=\columnwidth]{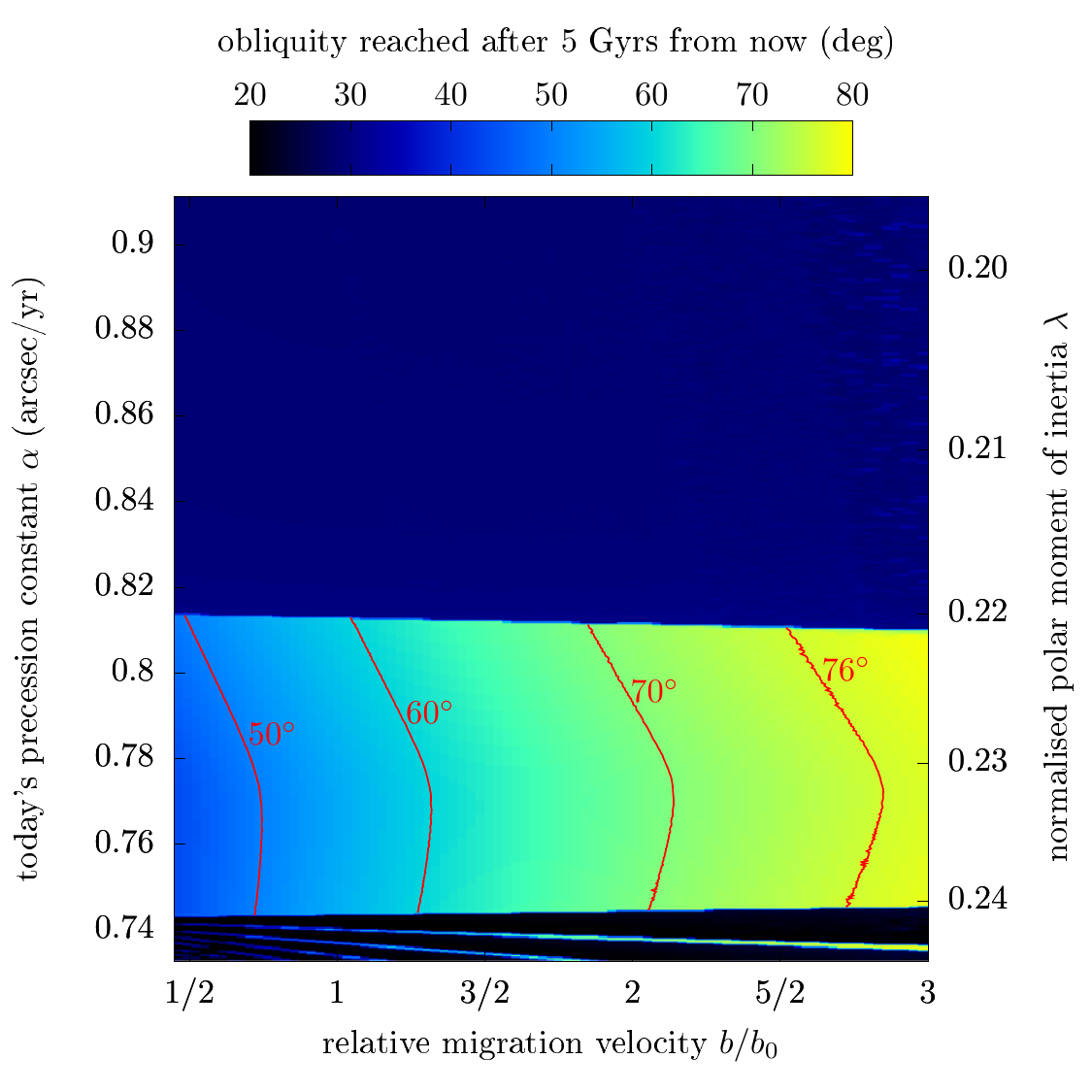}
      \caption{Future obliquity of Saturn as a function of Titan's migration velocity and Saturn's polar moment of inertia. The axes are the same as in Fig.~\ref{fig:obmin}. The $3\sigma$ uncertainty ranges of \cite{Lainey-etal_2020} yield approximately $b/b_0\in[1/2,5]$ for the astrometric measurements and $b/b_0\in[1,3/2]$ for the radio-science experiments. Some level curves are shown in red.}
      \label{fig:obmax}
   \end{figure}

   Figure~\ref{fig:obmax} shows the summary of our results for finely sampled values of $\lambda$ and $b$. Contrary to Fig.~\ref{fig:obmin}, we restrict here our sampling to $b<3b_0$ because for larger migration rates, Titan goes beyond the Laplace radius during the integration timespan ($a_6\approx 40$~$R_\mathrm{eq}$) and the close-satellite approximation used in Eq.~\eqref{eq:J2tilde} is invalidated. Faster migration rates are anyway disfavoured by the $3\sigma$ uncertainty range of the radio-science experiments of \cite{Lainey-etal_2020}.
   
   The top portion of Fig.~\ref{fig:obmax} features trajectories of Type~1. Such trajectories are currently above the resonance with $\phi_3$ (see Fig.~\ref{fig:types}) and they go farther away from it as $\alpha$ continues to increase. The increase in $\alpha$ makes them cross the resonances with $\phi_{51}$, with $\phi_{14}$, and with $\phi_{15}$ (see Fig.~\ref{fig:widths} and Table~\ref{tab:zetashort}). Being very small, these resonances are crossed quickly and they do not produce noticeable obliquity variations in Fig.~\ref{fig:obmax}. This explains why the top portion of the figure is coloured almost uniformly with an obliquity value approximatively equal to today's. For $b\approx 3b_0$ (the fastest migration presented in Fig.~\ref{fig:obmax}), trajectories reach the lower fringe of the strong resonance with $\phi_4$ at the end of the integration, but they do not actually reach it.
   
   \begin{figure}
	  \centering
	   \includegraphics[width=\columnwidth]{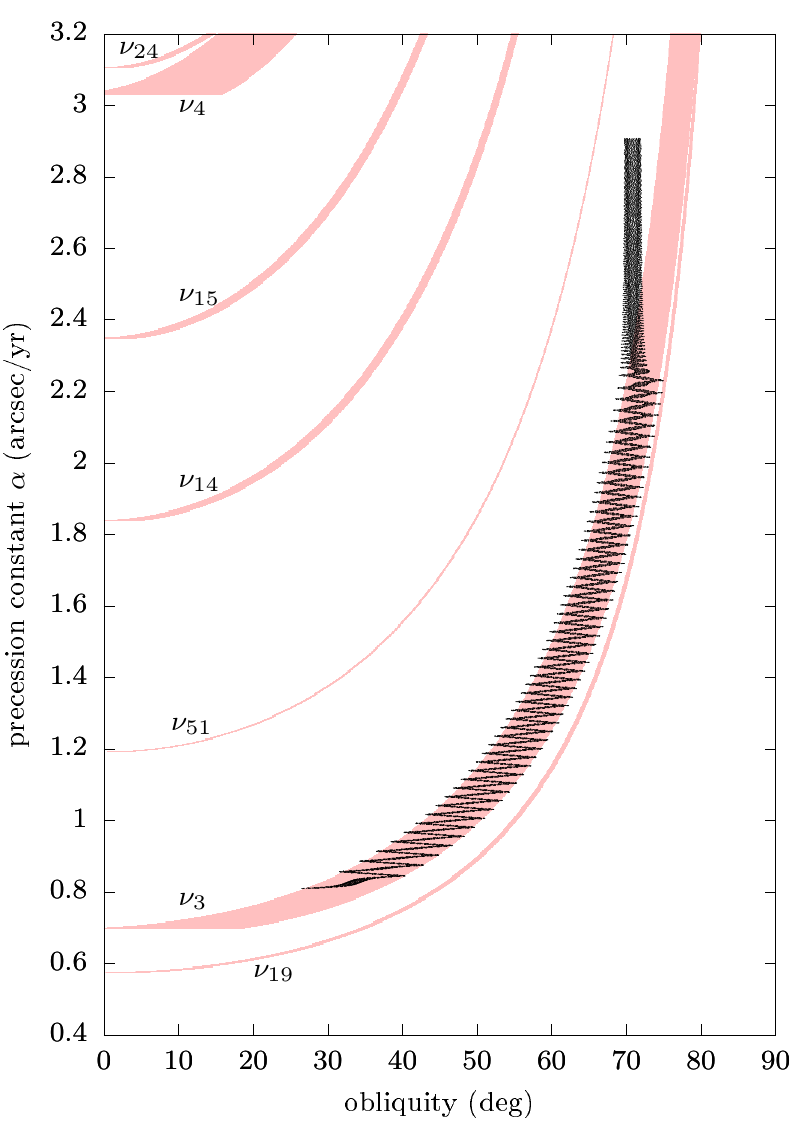}
	   \caption{Example of Type~2 trajectory that is expelled out of the resonance. It has been obtained for $\lambda=0.221076$ and a migration rate $b/b_0=3$. The integration runs from today up to $5$~Gyrs in the future.}
	   \label{fig:expul}
   \end{figure}
   
   The middle portion of Fig.~\ref{fig:obmax} features trajectories of Types~2 and~3. Such trajectories are currently inside the resonance with $\phi_3$ (see Fig.~\ref{fig:types}) and they follow its centre as $\alpha$ increases. After $5$~Gyrs from now, Saturn can therefore reach very large obliquity values, provided that Titan goes on migrating as predicted by \cite{Lainey-etal_2020}. For migration rates lying in the error range of the radio-science experiments of \cite{Lainey-etal_2020}, Saturn's obliquity can grow as large as $65^\circ$. As noticed by \cite{Saillenfest-etal_2020b}, the resonance width increases up to $\alpha\approx 0.971''\,$yr$^{-1}$, but decreases beyond. The trajectories featuring a large libration amplitude and a large increase in $\alpha$ have therefore a risk of being expelled out of the resonance, as described by \cite{Su-Lai_2020}. After a careful examination, we found that expulsion out of resonance only occurs for the largest migration velocities and in a tiny interval of $\lambda$ located at the very edge of the brightly-coloured region of Fig.~\ref{fig:obmax}. An example of such a trajectory is presented in Fig.~\ref{fig:expul}. The expelled trajectories reach slightly smaller values of obliquity than if they continued to follow the resonance centre; however, this behaviour concerns such a small range of parameters (which is almost undistinguishable in Fig.~\ref{fig:obmax}) that it is very unlikely to have any consequence for Saturn.
   
   The bottom portion of Fig.~\ref{fig:obmax} features trajectories of Type~4, which are ruled out by our uncertainty range for $\lambda\in[0.200,0.240]$. Such trajectories did not reach yet the resonance today (see Fig.~\ref{fig:types}), but they will in the future as $\alpha$ continues to increase. The resonance encounter can either lead to a capture (like Type~2 trajectories) or to a permanent decrease in obliquity (like Type~1 trajectories). The outcome is determined by the phase of $\sigma_3$ at crossing time, which depends on the migration velocity. This explains why the two possible outcomes are organised in Fig.~\ref{fig:obmax} as narrow bands that are close to each other for a slow migration and spaced for a fast migration. In a perfect adiabatic regime, the bands would be so close to each other that the outcome could be modelled as a probabilistic event.
   
\section{Discussion and conclusion}\label{sec:ccl}
   Since giant planets are expected to form with near-zero obliquities, some mechanism must have tilted Saturn after its formation. \cite{Saillenfest-etal_2020b} have shown that the fast migration of Titan measured by \cite{Lainey-etal_2020} may be responsible for the current large obliquity of Saturn. Through an extensive set of numerical simulations, we further investigated the long-term spin-axis dynamics of Saturn and determined the variety of laws for Titan's migration compatible with this scenario.
   
   Saturn is located today near a strong secular spin-orbit resonance with the nodal precession mode of Neptune \citep{Ward-Hamilton_2004}. As Titan migrates over time, it produces a drift in Saturn's spin-axis precession velocity, which led Saturn to encounter this resonance. The continuous migration of Titan shifts the resonance centre over time, which can force Saturn's obliquity to vary. Through this mechanism, Saturn's obliquity can have grown from a small to a large value provided that: \emph{i)} Titan migrated over a large enough distance to substantially shift the resonance centre, and \emph{ii)} Titan migrated slowly enough for Saturn to adiabatically follow the resonance shift. The first condition is met if Titan migrated over a distance of at least one radius of Saturn after the late planetary migration, more than $4$~Gyrs ago. Assuming that Titan's migration is continuous, this requires migration velocities larger than about $n\approx 0.06$~times the nominal rate given by \cite{Lainey-etal_2020}. For comparison, astrometric measurements predict $n\gtrsim 0.5$. The second condition is met if Titan's migration velocity does not exceed $n\approx 10$~times the nominal rate, while astrometric measurements predict $n\lesssim 5$. Therefore, the scenario proposed by \cite{Saillenfest-etal_2020b} is realistic over the whole range of migration rates obtained from observations. It even allows for more complex scenarios in which Titan would alternate between fast and slow migration regimes.
   
   For the largest migration rates of Titan allowed by observational uncertainty ranges, non-adiabatic effects are quite pronounced, but not to the point of preventing Saturn from following the resonance centre. Interestingly, non-adiabaticity even allows for an exactly zero value for Saturn's primordial obliquity. Zero values are however disfavoured by the error range of radio-science experiments, which yield most likely primordial obliquities between $2^\circ$ and $7^\circ$.
   
   Our Monte Carlo experiments do not reveal a strong chaotic mixing of trajectories, even though borderline separatrix-crossing trajectories do exhibit a noticeable chaotic spreading. All possible dynamical paths fall into the four types of trajectories obtained by \cite{Saillenfest-etal_2020b} through backward numerical integrations, and we detected no substantial numerical irreversibility. For Titan's nominal migration rate, our experiments show that all trajectories with initial obliquity smaller than about $10^\circ$ are captured inside the resonance with a $100\%$ probability. Such trajectories can match Saturn's current orientation if its normalised polar moment of inertia $\lambda$ lies in about $[0.224,0.237]$, as previously reported. Interestingly, small past obliquities $\varepsilon\lesssim 10^\circ$ in our Monte Carlo experiments also feature the highest likelihood of reproducing Saturn's current spin-axis orientation, surpassing high-obliquity alternatives by a factor of about ten. Yet, other values of $\lambda$ cannot be completely ruled out; they would mean that Saturn's past obliquity was larger or similar as today and one would need to find another explanation for its large value.
   
   In the future, the still ongoing migration of Titan is expected to produce dramatic effects on Saturn's obliquity provided that Saturn is currently located inside the resonance, that is, if $\lambda$ lies in about $[0.220,0.241]$. Depending on the precise migration rate of Titan, Saturn's obliquity would then range between $55^\circ$ and $65^\circ$ after $5$~Gyrs from now, and we even obtain values exceeding $75^\circ$ when considering the full $3\sigma$ uncertainty of the astrometric measurements of \cite{Lainey-etal_2020}. For smaller values of $\lambda$, Saturn's obliquity is not expected to change much in the future because the migration of Titan pushes it away from the resonance. No strong obliquity variations would be expected either if Titan's migration rate strongly drops in the future (i.e. if Titan is released out of the tidal resonance-locking mechanism of \citealp{Fuller-etal_2016}), but to our knowledge, there is no evidence showing that it could be the case.
   
   The migration law for Titan proposed by \cite{Lainey-etal_2020} and used in this article is very simplified. Since our conclusions remain valid in a much larger interval of migration rates than allowed by the observational uncertainties, we can be confident that no major change would be produced by using different (and possibly more realistic) migrations laws, unless Titan underwent extreme variations in migration rate in the past. For instance, if Titan's migration is not continuous and if it was only triggered very recently (less than a few hundreds of million years ago), then Saturn's past obliquity dynamics would not have been affected. As mentioned by \cite{Saillenfest-etal_2020b}, this alternative is unlikely but cannot be ruled out considering our current knowledge of the tidal dissipation within Saturn.
   
   The past and future behaviour of Saturn's spin axis is very sensitive to its normalised polar moment of inertia $\lambda$. An accuracy of at least three digits would be required to securely assert which dynamical path was followed by Saturn and what will be the future evolution of its spin axis. Model-dependent theoretical values are not enough for this purpose, and it is still unclear what is the true uncertainty of values inferred from the \emph{Cassini} data \citep{Helled_2011,Fortney-etal_2018,Movshovitz-etal_2020}. A precise value of $\lambda$ would inform us about whether Saturn is currently inside the resonance (which is the most likely alternative), or outside the resonance. If Saturn is confirmed to be currently in resonance, it would imply that Titan's past migration rate never became so fast as to eject Saturn from the resonance or to prevent its capture in the first place. However, this constraint would not be very stringent: simulations show that Saturn can be captured into resonance even if Titan's migration rate is increased by a factor ten from the nominal measured value. If, on the contrary, Saturn turns out to be currently out of resonance, then it would imply that its primordial obliquity was high, and most probably even higher than~$30^\circ$, regardless of Titan's precise migration history. This last possibility is not what one would expect from planetary formation models, and our results show that it is also unlikely in a dynamical point of view.
   
   Previous works reveal that numerous dynamical mechanisms can alter the obliquity of a planet (see e.g. \citealp{Laskar-Robutel_1993,Correia-Laskar_2001,Quillen-etal_2018,Millholland-Batygin_2019}). The fast migration of satellites and capture in a secular spin-orbit resonance offers one more alternative, and we have shown that it can result in a steady increase in obliquity, possibly lasting over the whole lifetime of the planetary system. In the broad context of exoplanets, we can therefore expect that only a few would have conserved their primordial axis tilt, whether they are close-in and likely tidally locked \citep{Millholland-Laughlin_2019}, or whether they are largely spaced and have very stable orbits like Jupiter and Saturn.
       
%
\begin{acknowledgements}
   Our work greatly benefited from discussions with David Nesvorn{\'y}; we thank him very much. We are also very grateful to Dan Tamayo for his in-depth review and inspiring comments. G.~L. acknowledges financial support from the Italian Space Agency (ASI) through agreement 2017-40-H.0.
\end{acknowledgements}
\bibliographystyle{aa}
\bibliography{saturnspin}
\appendix
\section{Orbital solution for Saturn}\label{asec:QPS}

The secular orbital solution of \cite{Laskar_1990} is obtained by multiplying the normalised proper modes $z_i^\bullet$ and $\zeta_i^\bullet$ (Tables VI and VII of \citealp{Laskar_1990}) by  the matrix $\tilde{S}$ corresponding to the linear part of the solution (Table V of \citealp{Laskar_1990}). In the series obtained, the terms with the same combination of frequencies are then merged together, resulting in 56 terms in eccentricity and 60 terms in inclination. This forms the secular part of the orbital solution of Saturn, which is what is required by our averaged model.

The orbital solution is expressed in the variables $z$ and $\zeta$ as described in Eqs.~\eqref{eq:qprep} and \eqref{eq:munu}. In Tables~\ref{tab:z} and \ref{tab:zeta}, we give all terms of the solution in the J2000 ecliptic and equinox reference frame.

\begin{table}
   \caption{Quasi-periodic decomposition of Saturn's eccentricity and longitude of perihelion (variable $z$).}
   \label{tab:z}
   \vspace{-0.7cm}
   \small
   \begin{equation*}
      \begin{array}{rrrr}
      \hline
      \hline
      k & \mu_k\ (''\,\text{yr}^{-1}) & E_k\times 10^8 & \theta_k^{(0)}\ (^\text{o}) \\
      \hline
       1 &  28.22069 & 4818642 & 128.11 \\
       2 &   4.24882 & 3314184 &  30.67 \\
       3 &  52.19257 &  173448 & 225.55 \\
       4 &   3.08952 &  151299 & 121.36 \\
       5 &  27.06140 &   55451 &  38.70 \\
       6 &  29.37998 &   54941 &  37.54 \\
       7 &  28.86795 &   32868 & 212.64 \\
       8 &  27.57346 &   28869 & 223.74 \\
       9 &  53.35188 &   14683 & 134.91 \\
      10 & -19.72306 &   14125 & 113.24 \\
      11 &  76.16447 &    7469 & 323.03 \\
      12 &   0.66708 &    5760 &  73.98 \\
      13 &   5.40817 &    4420 & 120.24 \\
      14 &  51.03334 &    4144 & 136.29 \\
      15 &   7.45592 &    1387 &  20.24 \\
      16 &   5.59644 &     805 & 290.35 \\
      17 &   1.93168 &     801 & 201.08 \\
      18 &   4.89647 &     717 & 291.46 \\
      19 &  17.36469 &     674 & 123.95 \\
      20 &   3.60029 &     408 & 121.39 \\
      21 &   2.97706 &     395 & 306.81 \\
      22 & -56.90922 &     365 &  44.11 \\
      23 &  17.91550 &     339 & 335.18 \\
      24 &   5.47449 &     303 &  95.01 \\
      25 &   5.71670 &     230 & 300.52 \\
      26 &  17.08266 &     187 & 179.38 \\
      27 & -20.88236 &     186 & 203.93 \\
      28 &   6.93423 &     167 & 349.39 \\
      29 &  16.81285 &     157 & 273.89 \\
      30 &   1.82121 &     139 & 151.70 \\
      31 &   7.05595 &     136 & 178.86 \\
      32 &   5.35823 &     124 & 274.88 \\
      33 &   7.34103 &     117 &  27.85 \\
      34 &   0.77840 &      99 &  65.10 \\
      35 &   7.57299 &      82 & 191.47 \\
      36 &  17.63081 &      78 & 191.55 \\
      37 &  19.01870 &      67 & 219.75 \\
      38 &  17.15752 &      64 & 325.02 \\
      39 &  17.81084 &      58 &  58.56 \\
      40 &  18.18553 &      53 &  57.27 \\
      41 &   5.99227 &      45 & 293.56 \\
      42 &  17.72293 &      44 &  48.46 \\
      43 &   5.65485 &      44 & 219.22 \\
      44 &   4.36906 &      39 &  40.82 \\
      45 &  16.52731 &      39 & 131.91 \\
      46 &   6.82468 &      38 &  14.53 \\
      47 &  18.01611 &      37 &  44.83 \\
      48 &   5.23841 &      36 &  92.97 \\
      49 &  17.47683 &      34 & 260.26 \\
      50 &  18.46794 &      32 &   4.67 \\
      51 &  -0.49216 &      29 & 164.74 \\
      52 &  17.55234 &      27 & 197.65 \\
      53 &  16.26122 &      26 &  58.89 \\
      54 &   7.20563 &      24 & 323.91 \\
      55 &  18.08627 &      22 & 356.17 \\
      56 &   7.71663 &      15 & 273.52 \\
      \hline
      \end{array}
   \end{equation*}
   \vspace{-0.5cm}
   \tablefoot{This solution has been directly obtained from \cite{Laskar_1990} as explained in the text. The phases $\theta_k^{(0)}$ are given at time J2000.}
\end{table}

\begin{table}
   \caption{Quasi-periodic decomposition of Saturn's inclination and longitude of ascending node (variable $\zeta$).}
   \label{tab:zeta}
   \vspace{-0.7cm}
   \small
   \begin{equation*}
      \begin{array}{rrrr}
      \hline
      \hline
      k & \nu_k\ (''\,\text{yr}^{-1}) & S_k\times 10^8 & \phi_k^{(0)}\ (^\text{o}) \\
      \hline
       1 &   0.00000 & 1377395 & 107.59 \\
       2 & -26.33023 &  785009 & 127.29 \\
       3 &  -0.69189 &   55969 &  23.96 \\
       4 &  -3.00557 &   39101 & 140.33 \\
       5 & -26.97744 &    5889 &  43.05 \\
       6 &  82.77163 &    3417 & 128.95 \\
       7 &  58.80017 &    2003 & 212.90 \\
       8 &  34.82788 &    1583 & 294.12 \\
       9 &  -5.61755 &    1373 & 168.70 \\
      10 & -17.74818 &    1269 & 123.28 \\
      11 & -27.48935 &    1014 & 218.53 \\
      12 & -25.17116 &     958 & 215.94 \\
      13 & -50.30212 &     943 & 209.84 \\
      14 &  -1.84625 &     943 &  35.32 \\
      15 &  -2.35835 &     825 & 225.04 \\
      16 &  -4.16482 &     756 &  51.51 \\
      17 &  -7.07963 &     668 & 273.79 \\
      18 & -28.13656 &     637 & 314.07 \\
      19 &  -0.58033 &     544 &  17.32 \\
      20 &  -5.50098 &     490 & 162.89 \\
      21 &  -6.84091 &     375 & 106.28 \\
      22 &  -7.19493 &     333 & 105.15 \\
      23 &  -6.96094 &     316 &  97.96 \\
      24 &  -3.11725 &     261 & 326.97 \\
      25 &  -7.33264 &     206 & 196.75 \\
      26 & -18.85115 &     168 &  60.48 \\
      27 &  -5.85017 &     166 & 345.47 \\
      28 &   0.46547 &     157 & 286.88 \\
      29 & -19.40256 &     141 & 208.18 \\
      30 & -17.19656 &     135 & 333.96 \\
      31 &  -5.21610 &     124 & 198.91 \\
      32 &  -5.37178 &     123 & 215.48 \\
      33 &  -5.10025 &     121 &  15.38 \\
      34 & -18.01114 &      96 & 242.09 \\
      35 & -17.66094 &      91 & 138.93 \\
      36 &  11.50319 &      83 & 281.01 \\
      37 & -17.83857 &      74 & 289.13 \\
      38 &  -5.96899 &      71 & 170.64 \\
      39 &  -6.73842 &      67 &  44.50 \\
      40 & -17.54636 &      66 & 246.71 \\
      41 &  -7.40536 &      62 & 233.35 \\
      42 &  -7.48780 &      58 &  47.95 \\
      43 &  -6.56016 &      54 & 303.47 \\
      44 &   0.57829 &      54 & 103.72 \\
      45 &  -6.15490 &      51 & 269.77 \\
      46 & -17.94404 &      47 & 212.26 \\
      47 &  -8.42342 &      45 & 211.21 \\
      48 & -18.59563 &      43 &  98.11 \\
      49 &  20.96631 &      32 &  57.78 \\
      50 &   9.18847 &      31 &   1.15 \\
      51 &  -1.19906 &      30 & 132.74 \\
      52 &  10.34389 &      20 & 190.42 \\
      53 &  18.14984 &      19 & 291.19 \\
      54 & -19.13075 &      18 & 305.90 \\
      55 & -18.97001 &       8 &  73.36 \\
      56 & -18.30007 &       7 & 250.45 \\
      57 & -18.69743 &       4 & 221.70 \\
      58 & -18.77933 &       4 & 222.83 \\
      59 & -18.22681 &       4 &  46.30 \\
      60 & -19.06544 &       4 &  50.21 \\
      \hline
      \end{array}
   \end{equation*}
   \vspace{-0.5cm}
   \tablefoot{This solution has been directly obtained from \cite{Laskar_1990} as explained in the text. The phases $\phi_k^{(0)}$ are given at time J2000.}
\end{table}

\section{Examples of trajectories featuring extreme phase effects}\label{asec:extreme}

In Sect.~\ref{ssec:extreme}, we show that trajectories crossing the separatrix can feature extreme phase effects when they reach the resonance in the vicinity of its hyperbolic point and follow its drift over time. This maintains them on the edge between capture (Type~2 trajectory) and non-capture (Type~1 trajectory).

\begin{figure*}
   \centering
   \includegraphics[width=0.99\columnwidth]{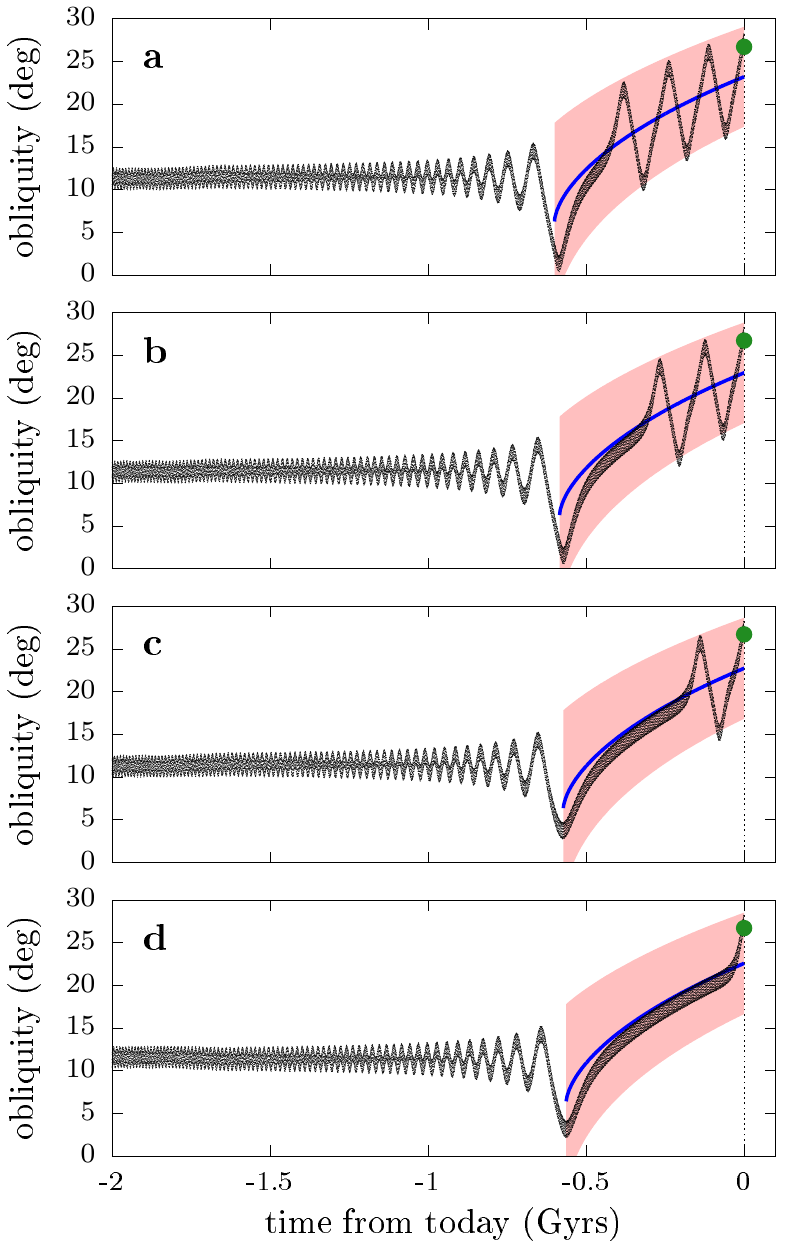}
   \includegraphics[width=0.99\columnwidth]{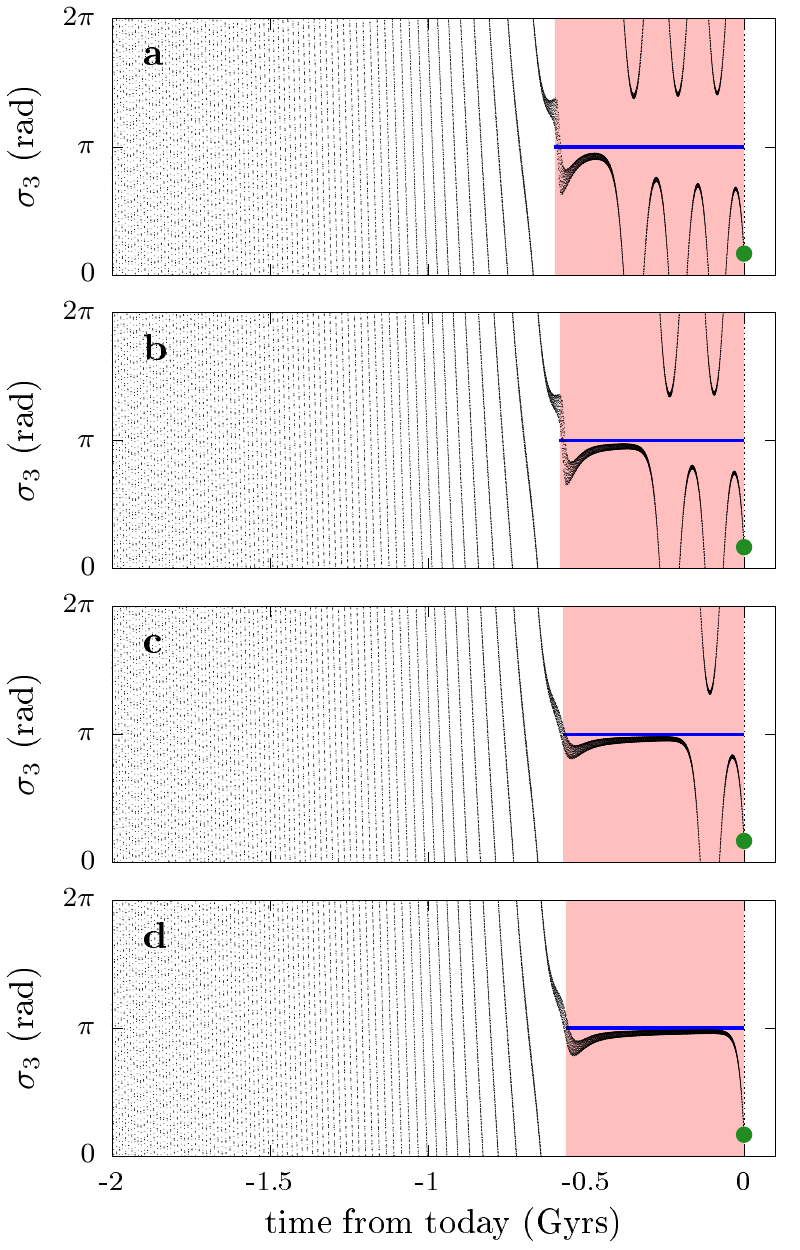}
   \caption{Example of trajectories featuring an extreme phase effect. The left column shows the evolution of the obliquity, and the right column shows the evolution of the resonant angle $\sigma_3=\psi+\phi_3$. The migration parameter is $b=b_0$. For each row, the parameter $\lambda$ used corresponds to the tip of a spike in Fig.~\ref{fig:phase_effects} (see labels), tuned at the $10^{-15}$ level. The pink area represents the interval occupied by the resonance once the separatrix appears. The blue curve shows the location of the hyperbolic equilibrium point (Cassini state 4). The green point shows Saturn current location (at $t=0$).}
   \label{fig:extreme}
   \includegraphics[width=0.99\columnwidth]{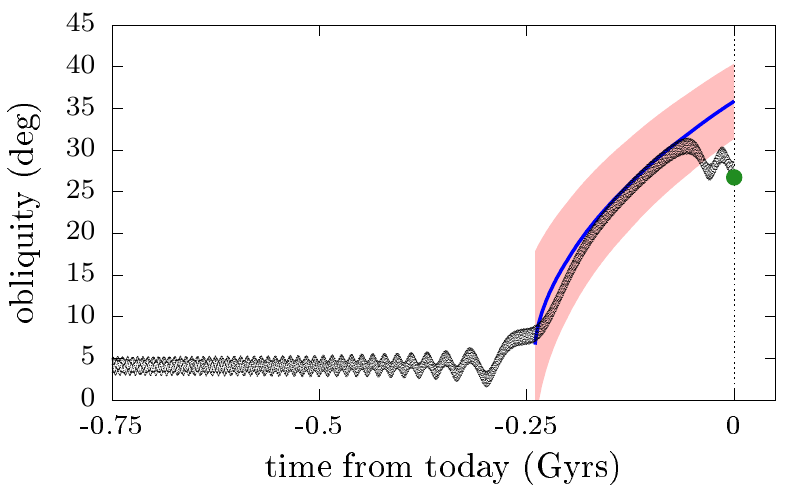}
   \includegraphics[width=0.99\columnwidth]{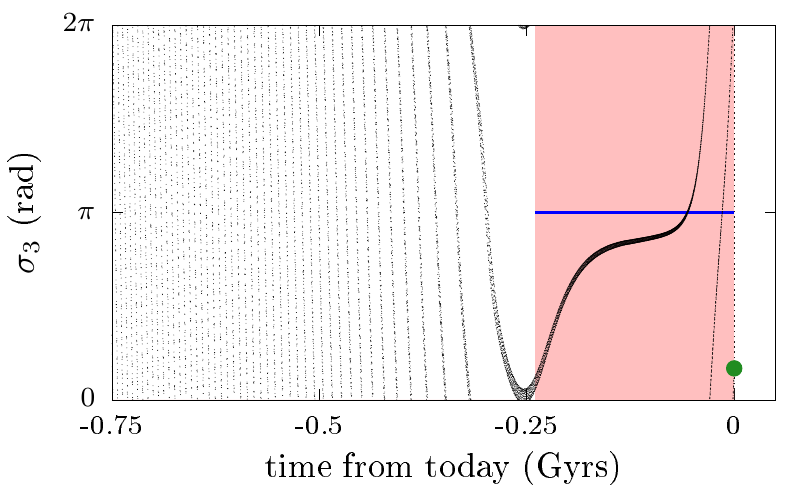}
   \caption{Same as Fig.~\ref{fig:extreme}, but for a trajectory of Type~1. This trajectory has a migration parameter $b=7.37\,b_0$ and a normalised polar moment of inertia $\lambda=0.2114$.}
   \label{fig:extremeT1}
\end{figure*}

Figure~\ref{fig:extreme} shows examples of such trajectories obtained for Titan's nominal migration rate. These trajectories are of Type~2 (i.e. currently inside the resonance). Instead of the precession angle $\psi$, we plot the resonant angle $\sigma_3 = \psi+\phi_3$, where $\phi_3$ evolves as in Eq.~\eqref{eq:munu}. The elliptic point of the resonance (Cassini state 2) is located at $\sigma_3=0$, and the hyperbolic equilibrium point (Cassini state 4) is located at $\sigma_3=\pi$ (see e.g. \citealp{Saillenfest-etal_2019a}). We see that passing from one spike of Fig.~\ref{fig:phase_effects} to the next one corresponds to performing one more oscillation inside the resonance. For a purely adiabatic dynamics, all spikes would be infinitely close to each other, such that it would be impossible to get one specific trajectory by finely tuning $\lambda$.

Figure~\ref{fig:extremeT1} shows another example of extreme phase effect but for a trajectory of Type~1 (i.e. currently outside the resonance). It is obtained using a strongly non-adiabatic migration, which widens the parameter ranges allowing for extreme phase effects (see Sect.~\ref{ssec:extreme}). This trajectory does not exactly match Saturn's spin-axis orientation today, but it lies within our strict success criteria defined in Sect.~\ref{ssec:MCstrict}: its current coordinates $\varepsilon$ and $\psi$ are within $0.4^\circ$ and $4.9^\circ$ of the actual ones, respectively. This trajectory appears in the top portion of Fig.~\ref{fig:constraint} as the leftmost isolated grey point. It can be linked to the bottom of a spike in Fig.~\ref{fig:slowfast}, that is, to one of the top blue stripes of Fig.~\ref{fig:obmin}. After having bifurcated away from the hyperbolic point, Fig.~\ref{fig:extremeT1} shows that this trajectory has performed one complete revolution of $\sigma_3$. The two other isolated grey points in Fig.~\ref{fig:constraint} have performed zero and two, respectively.

\section{Experiments on the initial obliquity prior}\label{ssec:prior}

In Sect.~\ref{ssec:MCstrict}, a Monte Carlo experiment is performed in order to look for the most likely values of Saturn's precession constant and Titan's migration rate. Formation models predict that Saturn's primordial obliquity was near-zero, but the statistics obtained greatly depend on the precise distribution used as initial conditions. In this section, we investigate further this dependence with additional Monte Carlo experiments.

\begin{figure*}
   \centering
   \includegraphics[width=\textwidth]{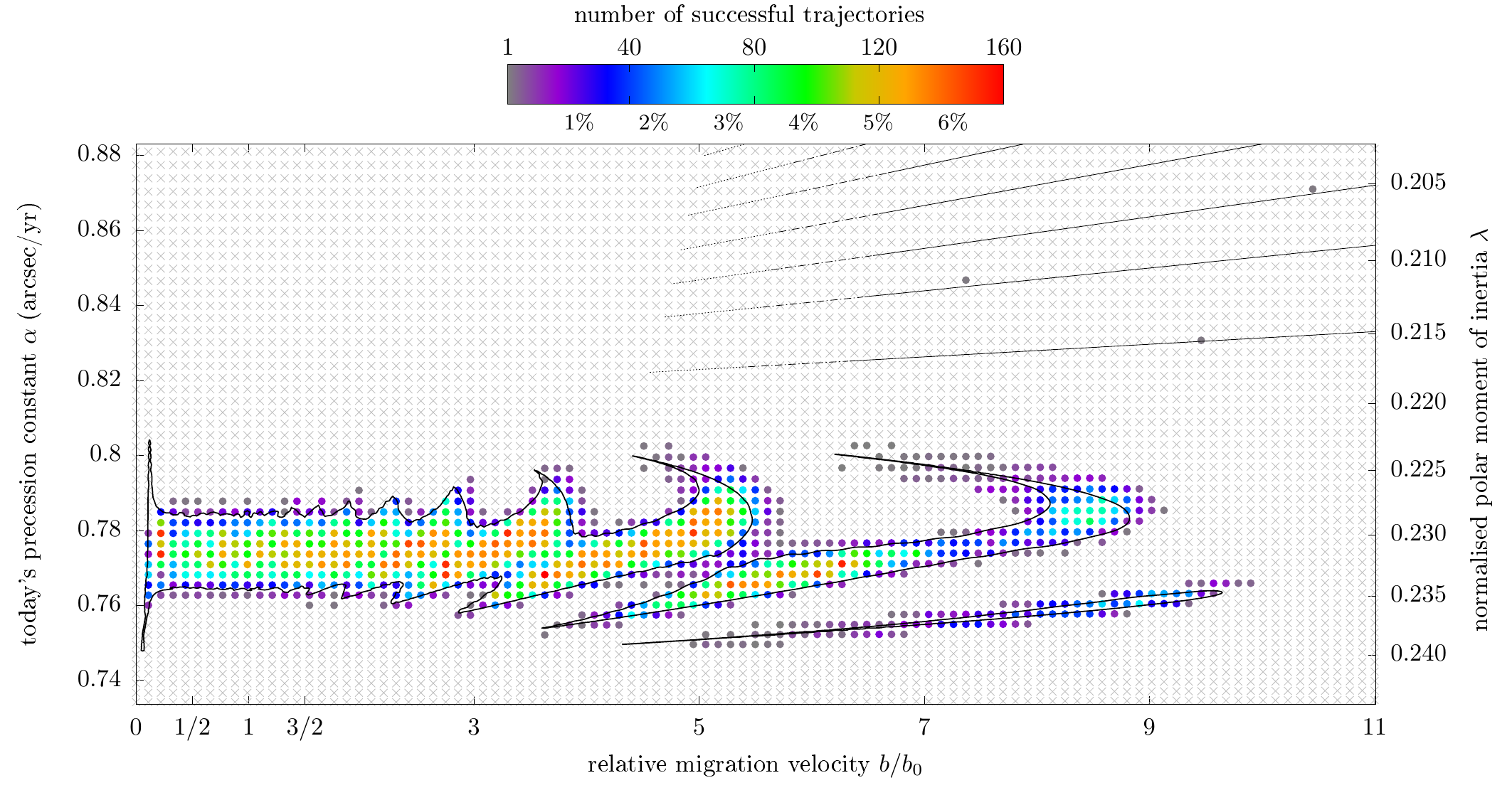}
   \caption{Same as Fig.~\ref{fig:constraint}, but considering an isotropic distribution of initial spin orientation with $\varepsilon\leqslant 5^\circ$. It is obtained from Fig.~\ref{fig:constraint} by weighting the count of each run by $\sin\varepsilon$ (see Fig.~\ref{fig:histog}).}
   \label{fig:constraint_weight}
\end{figure*}

\begin{figure}
   \centering
   \includegraphics[width=\columnwidth]{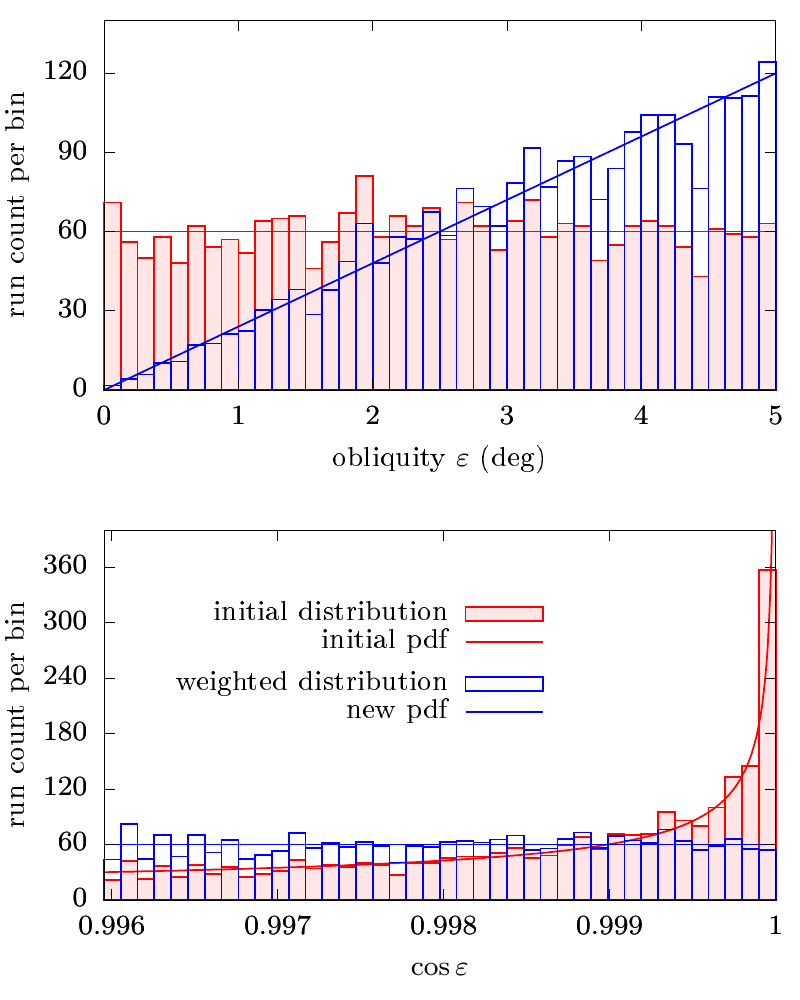}
   \caption{Sampled distribution of initial obliquity for one arbitrary point of Fig.~\ref{fig:constraint}, made of $2400$ simulations. The raw count of sampled trajectories is shown in red; the weighted count is shown in blue. Top: histogram with respect to the obliquity. Bottom: histogram with respect to the cosine of obliquity. The probability density functions (`pdf') are shown by the red and blue curves.}
   \label{fig:histog}
\end{figure}

Figure~\ref{fig:constraint_weight} shows the statistics obtained when assuming a uniform distribution of initial conditions over the spherical cap defined by $\varepsilon\leqslant 5^\circ$ (i.e. with a uniform sampling of $\cos\varepsilon$ instead of $\varepsilon$). Contrary to Fig.~\ref{fig:constraint}, this distribution is isotropic: it assumes that all directions over the spherical cap are equiprobable; small obliquity values are not particularly favoured. In practice, we can avoid running millions of simulations again by simply weighting the count of each run in Fig.~\ref{fig:constraint} by $\sin\varepsilon$. As illustrated in Fig.~\ref{fig:histog}, this trick allows us to mimic a uniform distribution of $\cos\varepsilon$ from a uniform distribution of $\varepsilon$. This method has the drawback of reducing by roughly a factor two the resolution at the high-obliquity end of the distribution (since trajectories are weighted by a factor $w\approx 2$), but this is not a problem here thanks to the high number of simulations.

\begin{figure*}
   \centering
   \includegraphics[width=\textwidth]{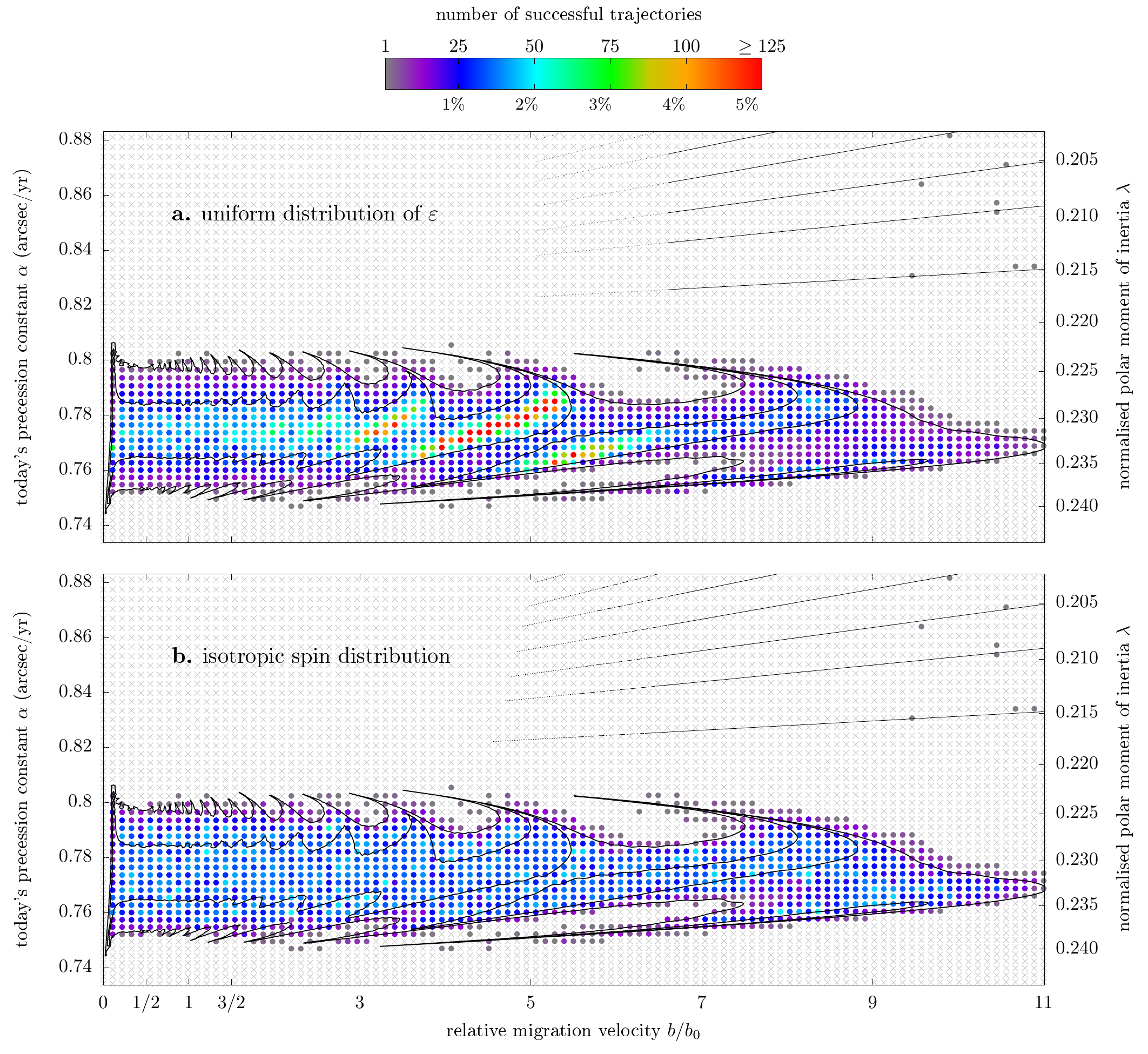}
   \caption{Same as Fig.~\ref{fig:constraint}, but for a range of initial spin orientations enlarged to $0^\circ\leqslant\varepsilon\leqslant 10^\circ$. Each point is made of $2400$ numerical simulations. \textbf{a}: uniform random distribution of $(\varepsilon,\psi)\in[0^\circ,10^\circ]\times[0,2\pi)$. \textbf{b}: uniform random distribution of $(\varepsilon,\psi)$ over the spherical cap defined by $\varepsilon\leqslant 10^\circ$. As in Fig.~\ref{fig:constraint_weight}, Panel~\textbf{b} is obtained by weighting the count numbers of Panel~\textbf{a}. The black contours show the $5^\circ$ and $10^\circ$ levels obtained through backward numerical integrations (see Fig.~\ref{fig:obmin}).}
   \label{fig:constraint10}
\end{figure*}

Interestingly, Fig.~\ref{fig:constraint_weight} shows that a uniform distribution of initial spin directions over the sphere yields approximatively equal likelihoods for the adiabatic ($b\lesssim 3b_0$) and non-adiabatic ($b\gtrsim 3b_0$) regimes. If we enlarge the distribution of initial conditions to $\varepsilon\leqslant 10^\circ$, Fig.~\ref{fig:constraint10} shows that the limits between the adiabatic and non-adiabatic regimes completely vanish, leaving only one large region with roughly constant likelihood.

\begin{figure*}
   \centering
   \includegraphics[width=\textwidth]{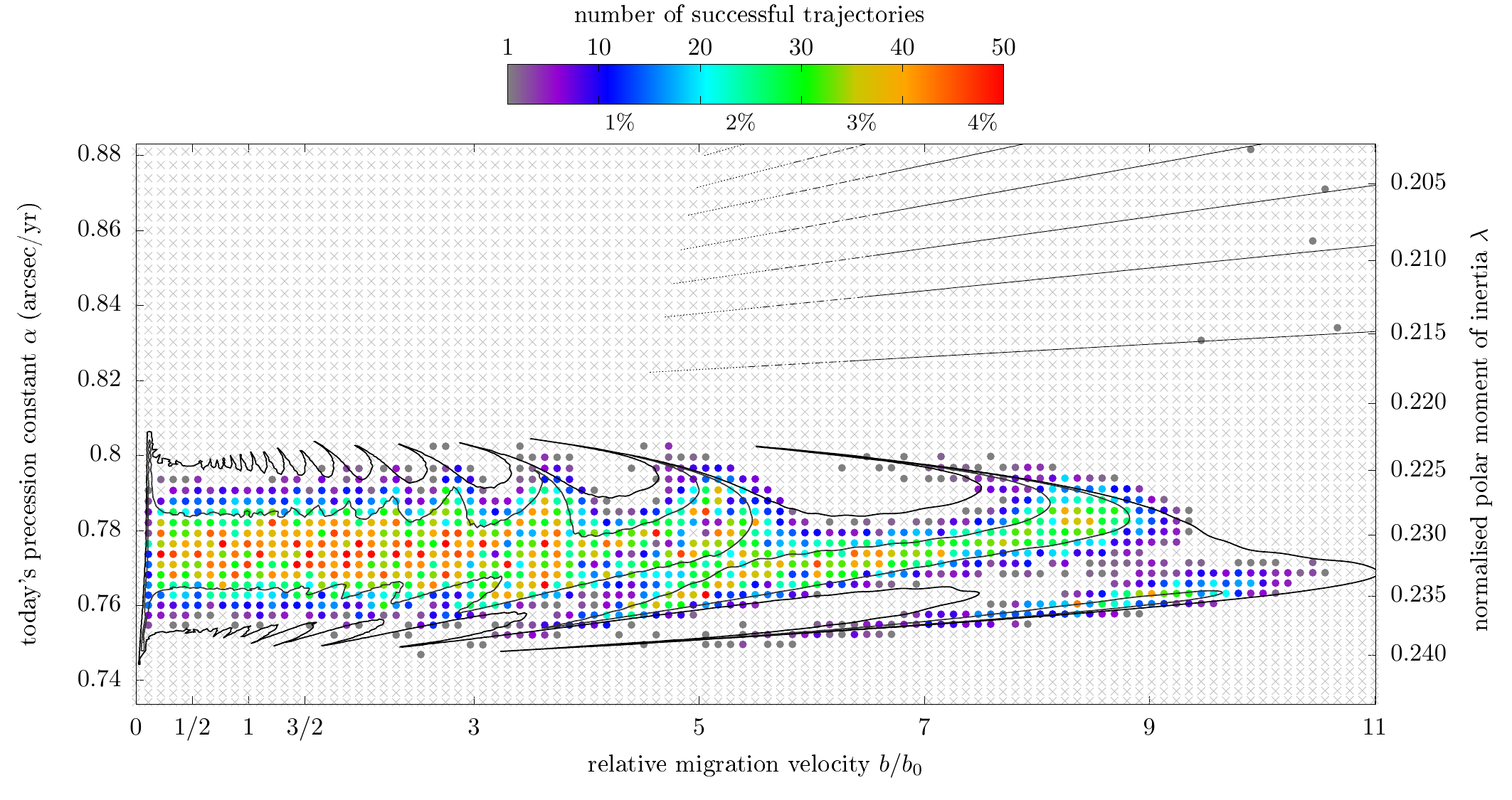}
   \caption{Same as Fig.~\ref{fig:constraint}, but for initial obliquities uniformly distributed between $2.5^\circ$ and $7.5^\circ$. It is obtained from a sub-sample of Fig.~\ref{fig:constraint10}a, such that each point is made of about $1200$ runs. The black contours show the $5^\circ$ and $10^\circ$ levels obtained through backward numerical integrations (same as Fig.~\ref{fig:constraint10}).}
   \label{fig:constraint27}
\end{figure*}

Figure~\ref{fig:constraint27} shows the distribution of successful runs starting from initial obliquities in the range $2.5^\circ\leqslant\varepsilon\leqslant 7.5^\circ$. This interval turns out to be roughly the one that favours most the adiabatic regime and Titan's nominal migration rate, to the detriment of the non-adiabatic regime. This is not surprising, since past obliquities between about $2^\circ$ and $7^\circ$ are the most likely for Titan's nominal migration rate (see Fig.~\ref{fig:CIsearch_strong}).

\end{document}